\documentclass[10pt,journal,compsoc]{IEEEtran}
\ifCLASSOPTIONcompsoc
  \usepackage[nocompress]{cite}
\else
  \usepackage{cite}
\fi
\ifCLASSINFOpdf
   \usepackage[pdftex]{graphicx}
   \graphicspath{{../pdf/}{../jpeg/}}     
    \DeclareGraphicsExtensions{.pdf,.jpeg,.png}
\else
    \usepackage[dvips]{graphicx}
    \graphicspath{{../eps/}}
    \DeclareGraphicsExtensions{.eps}
\fi

\usepackage{hyperref}
\usepackage{booktabs}
\usepackage{cite}
\usepackage{amsmath}
\usepackage{algorithm}
\usepackage{color} 
\makeatletter
\newcommand{\removelatexerror}{\let\@latex@error\@gobble}
\makeatother
\usepackage{amsthm,amsmath,amssymb}
\usepackage{makecell}
\usepackage{siunitx} 
\usepackage{amsfonts}
\usepackage{mathrsfs}
\usepackage{indentfirst}
\usepackage{multirow}
\usepackage{cleveref}
\usepackage{hyperref}

\usepackage{algorithmic}
\ifCLASSOPTIONcompsoc
  \usepackage[caption=false,font=footnotesize,labelfont=sf,textfont=sf]{subfig}
\else
  \usepackage[caption=false,font=footnotesize]{subfig}
\fi
\usepackage{fixltx2e}

\usepackage[T1]{fontenc}
\usepackage[utf8]{inputenc}
\usepackage{dblfloatfix}

\begin{document}
%
\title{Semi-WTC: A Practical Semi-supervised Framework for Attack Categorization through Weight-Task Consistency}
%
%
%
%

\author{Zihan~Li*,
        ~Wentao~Chen*,
        ~Zhiqing~Wei,~\IEEEmembership{Member,~IEEE},%
        ~Xingqi~Luo,
        and~Bing~Su\textsuperscript{\dag}
\IEEEcompsocitemizethanks{\IEEEcompsocthanksitem Z. Li is with the School of Informatics, Xiamen University, Xiamen 361005, China. E-mail: \href{zihanli@stu.xmu.edu.cn}{zihanli@stu.xmu.edu.cn}

\IEEEcompsocthanksitem W. Chen is with the School of Information and Communication Engineering, Beijing University of Posts and Telecommunications, Beijing 100876, China. E-mail: \href{wentaochen@bupt.edu.cn}{wentaochen@bupt.edu.cn}

\IEEEcompsocthanksitem Z. Wei is with the Key Laboratory of Universal Wireless Communications, Ministry of Education, Beijing University of Posts and Telecommunications, Beijing 100876, China. E-mail: \href{weizhiqing@bupt.edu.cn}{weizhiqing@bupt.edu.cn}

\IEEEcompsocthanksitem X. Luo is with the School of Computer Science \& Technology, Beijing Institute of Technology, Beijing 100081, China. E-mail: \href{1120182901@bit.edu.cn}{1120182901@bit.edu.cn}

\IEEEcompsocthanksitem B. Su is with the Beijing Key Laboratory of Big Data Management and Analysis Methods, Gaoling School of Artificial Intelligence, Renmin University of China, Beijing 100872, China. E-mail: \href{bingsu@ruc.edu.cn}{bingsu@ruc.edu.cn}
}
\thanks{ Equal contributions(*),  Corresponding author ({$\dag$}).}
}

\markboth{IEEE TRANSACTIONS ON DEPENDABLE AND SECURE COMPUTING,~August~2022}
{Shell \MakeLowercase{\textit{et al.}}: Bare Demo of IEEEtran.cls for Computer Society Journals}
%


\IEEEtitleabstractindextext{\begin{abstract}
Supervised learning has been widely used for attack categorization, requiring high-quality data and labels. However, the data is often imbalanced and it is difficult to obtain sufficient annotations. Moreover, supervised models are subject to real-world deployment issues, such as defending against unseen artificial attacks. To tackle the challenges, we propose a semi-supervised fine-grained attack categorization framework consisting of an encoder and a two-branch structure and this framework can be generalized to different supervised models. The multilayer perceptron with residual connection is used as the encoder to extract features and reduce the complexity. The Recurrent Prototype Module (RPM) is proposed to train the encoder effectively in a semi-supervised manner. To alleviate the data imbalance problem, we introduce the Weight-Task Consistency (WTC) into the iterative process of RPM by assigning larger weights to classes with fewer samples in the loss function. In addition, to cope with new attacks in real-world deployment, we propose an Active Adaption Resampling (AAR) method, which can better discover the distribution of unseen sample data and adapt the parameters of encoder. Experimental results show that our model outperforms the state-of-the-art semi-supervised attack detection methods with a 3\% improvement in classification accuracy and a 90\% reduction in training time.
\end{abstract}

}

\maketitle
\IEEEdisplaynontitleabstractindextext
%
\IEEEpeerreviewmaketitle
\IEEEraisesectionheading{\section{Introduction}\label{sec:introduction}}
\IEEEPARstart{O}{UR} lives are closely related to the Internet. As network technology evolves, network security issues have become more critical. Network traffic anomalies can degrade the network's performance and even make the network unavailable. It is mainly due to network attacks, misconfigurations, and abnormal network interruptions. There are many kinds of common network attacks, such as distributed denial of service (DDoS) attacks and brute force. Network traffic anomaly detection, especially identifying fine-grained anomalies or attack categories, can provide important support for network security. Network anomaly detection models \cite{52} are first proposed in host intrusion detection. Among them, snort \cite{53} and suricata \cite{54} intrusion detection systems are based on rule matching algorithms, which cannot detect unseen attacks. Nowadays, machine learning-based network intrusion detection algorithms \cite{50}\cite{51} that can detect unseen attacks are flourishing. Existing methods can be divided into two categories: coarse-grained attack detection \cite{55}\cite{56}\cite{57}\cite{58} and fine-grained attack detection\cite{12}\cite{15}. Coarse-grained methods only distinguish between attack and normal, which are not well suited for detailed analysis of attacks. Fine-grained methods further classify attack types to better respond to network attacks, so they are the focus of our research. Implementing fine-grained attack detection has the following challenges:

\textbf{(i) The difficulty of obtaining high-quality data and labels.} The categories within existing NetFlow datasets are often unbalanced. For example, in the NSL-KDD dataset, the number of samples in a head category may be hundreds of times larger than the number of samples in a tail category. Training directly on highly imbalanced data usually over-fits the dominant classes. In addition, due to the high cost of labeling, we have limited access to labeled data. Some attack samples have only coarse-grained labels and suffer from missing fine-grained labels. Also the various sample features from imbalanced data will cause the uncertainty of the model during the training process. Semi-supervised learning \cite{23} and self-supervised learning \cite{2} that can learn from unlabeled data have received increasing attention because they can reduce the cost of data labeling. Many recent approaches \cite{2}\cite{8}\cite{22} merge unlabeled data by performing consistency regularization to address the problem of class imbalance. 

In the meantime, deep learning and semi-supervised learning have made significant progress in computer vision and image segmentation \cite{74}\cite{75}. The representative ones are the DTC model and the CPS model. The DTC model \cite{8} divides the whole model into supervised and unsupervised parts and achieves contextual consistency in a pixel-to-pixel manner using directional contrast loss (DC Loss). The CPS model \cite{21} uses two parallel segmentation networks for cross pseudo supervision. These two models have inspired us. However, there is often an imbalance in the data distribution, which these two methods cannot solve, and few works have studied the coexistence of both category imbalance and insufficient data labeling.

\textbf{(ii) The lack of representative sample selection.} Most previous semi-supervised models use a small fraction of the labels in a given dataset to learn features \cite{van2020survey}. There lacks an effective mechanism to actively pick data representative of the distribution from all labeled data. Properly selecting such informative samples for labeling allows the model to generalize better.

\textbf{(iii) Multiple practical problems in real-life deployments.} \textbf{1)} The model cannot be too complex to reduce the detection efficiency. \textbf{2)} Artificial attacks (such as adversarial attacks and poisoning attacks) affect the usability of the model. \textbf{3)} Different intrusion detection models extract different features, making the paradigm not universal and cannot be extended to other security domains. Regarding the problem of artificial attacks, existing methods only study coarse-grained attacks \cite{1} and cannot distinguish fine-grained attacks. Most intrusion detection methods have customized features for feature extraction and are not generic \cite{4}.

\begin{figure}[!ht]
    \setlength{\belowcaptionskip}{0pt}
	\centerline{\includegraphics[width=0.45\textwidth]{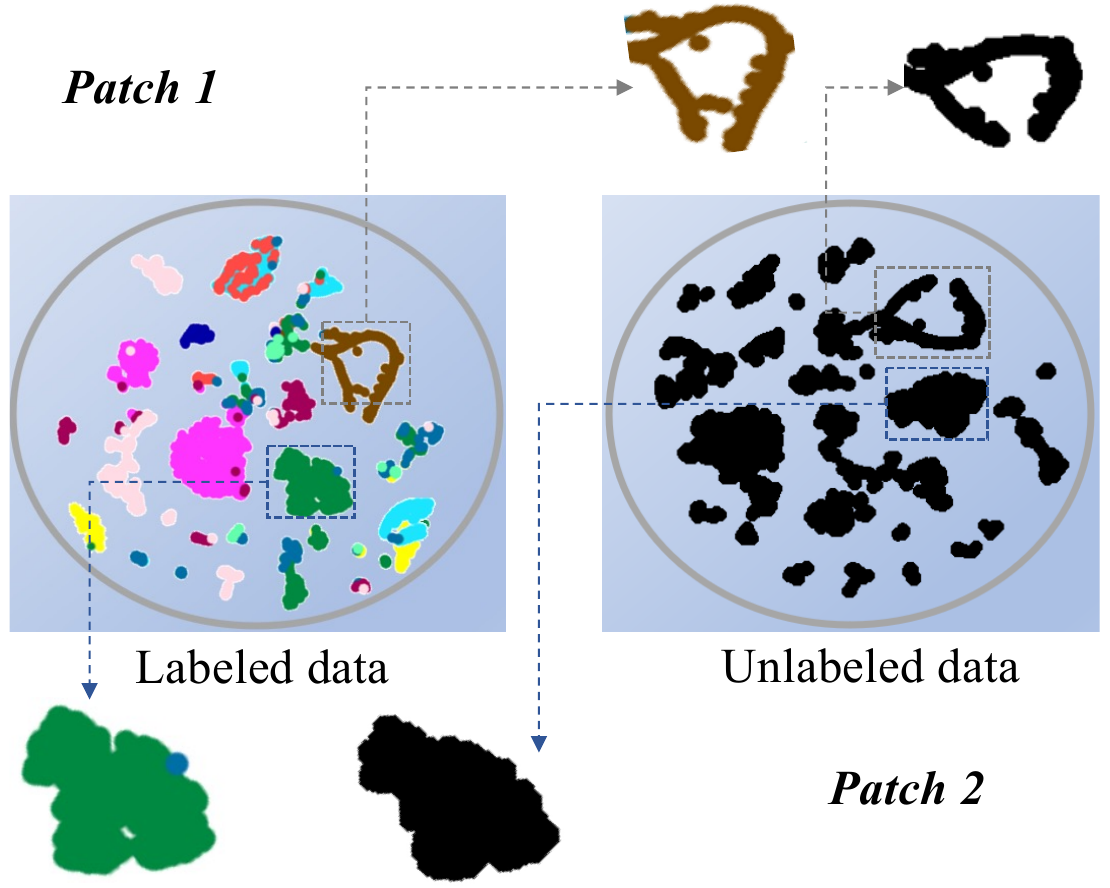}}
	\caption{Distribution of unlabeled and labeled data on the NSL-KDD dataset. The label distribution of the dataset is imbalanced. It is mainly reflected that a few classes (head classes) occupy most of the dataset. In contrast, most classes (tail classes) have tiny data samples.}
	\label{f1}
	\vspace{-2mm}
\end{figure}

In addition to the above challenges, Our work is also motivated by the observation of data in Fig. \ref{f1}. To address the effects of unlabeled and unknown data classes, we first observe the characteristics of the distribution of unlabeled and labeled data in the dataset. Then, taking the NSL-KDD dataset as an example, we randomly remove some sample labels to artificially divide the rest into a labeled dataset and an unlabeled dataset. We find that the data distribution of the unlabeled data is highly consistent with that of the labeled data, as shown in Fig. \ref{f1}. This distribution indicates that the feature information within the unlabelled data is not different from that within the labeled data itself. Inspired by this, discovering the potential information in the unlabeled data can improve the generalization ability of the network.

Together with the motivation, our research focuses on how to break through the performance bottlenecks of insufficient labeling (\textbf{Challenges i}, \textbf{ii}) and on how to overcome the practical bottlenecks of actually deploying the model (\textbf{Challenge iii}). To tackle the above challenges, we propose a practical intrusion detection framework. Concerning \textbf{Challenge (i)} and motivation of discovering the potential information in the unlabeled data, we propose a recurrent prototype module (RPM) to classify fine-grained attacks in a semi-supervised manner to utilize both labeled and unlabeled data simultaneously. We further incorporate the Weight-Task Consistency (WTC) to compensate for the imbalanced data, as well as Class Uncertainty Module (CUM) to reduce the uncertainty during training caused by data imbalance. Concerning \textbf{Challenge (ii)}, we propose an active adaption resampling (AAR) method to select representative samples. Concerning \textbf{Challenge (iii)}, we employ a simple residual connection architecture (RB-MLP) to reduce the complexity, taking any extracted features as input for further feature transformation and, hence, having better generality. We have implemented the prototype system\footnote{\href{https://github.com/HUANGLIZI/WTC}{https://github.com/HUANGLIZI/WTC}.}. The main contributions of the paper are summarized as follows.

\begin{enumerate}
\item We propose a practical semi-supervised framework (Semi-WTC) for fine-grained attack detection. To the best of our knowledge, Semi-WTC is the first perceptron-based framework for consistent semi-supervised intrusion detection and the first semi-supervised attack categorization model that considers both long-tail distributions and unseen classes. Experiments indicate that Semi-WTC achieves state-of-the-art performance.

\item We introduce a WTC module into the RPM iterative process, minimizing the difference between the prediction results from the unsupervised and supervised branches. It re-balances data of different fine-grained classes and speeds up the model convergence. Moreover, it can be generalized to the other attack detection models.

\item To handle unseen attacks, we propose the Active Adaption Resampling (AAR) method to label unseen samples that are more representative and improve the model's generalization ability. 
To alleviate the problem of model uncertainty in the long-tail distribution, we propose the Class Uncertainty Module (CUM) to suppress irrelevant features which cause the uncertainty.
\end{enumerate}

\vspace{-4mm}
\section{Related Work}

\subsection{\textbf{Machine Learning-based Intrusion Detection}} 
Machine learning-based attack detection can be divided 
into coarse-grained attack detection \cite{55}\cite{56} and fine-grained attack detection \cite{12}\cite{15}. DiFF-RF \cite{55} introduced an ensemble approach composed of random partitioning binary trees to detect anomalies, while Whisper \cite{56} introduced a real-time machine learning-based malicious traffic detection system thorough utilizing sequential information represented by the frequency domain features to achieve bounded information loss, as well as ensuring high detection accuracy. In cyber security, FARE \cite{12} utilizes eans clustering as an unsupervised classification algorithm supplemented with partially labeled data for semi-supervised learning. Recently, Diallo \cite{15} proposed an intrusion detection model based on adaptive clustering (ACID), which uses low-dimensional embeddings learned by efficient convolutional neural networks to solve the problem that traditional classification models are sensitive to feature changes. ACID can effectively separate samples from different classes. The existing fine-grained attacks, such as adversarial attacks, have not been sufficiently implemented yet. Therefore, we can consider machine learning methods to solve the problem of intrusion detection.

\vspace{-4mm}
\subsection{\textbf{Semi-supervised Learning}} 
We discuss semi-supervised learning methods in related fields. Semi-supervised learning methods can autonomously update the underlying machine learning models and be deployed in real-world environments. Effective semi-supervised methods for fine-grained attack detection are relatively lacking \cite{r2xu2020method}. MixMatch \cite{44} introduced an incremental broadening-based approach to implement semi-supervision for image classification. S4l \cite{68} introduced a framework unifying semi-supervised learning and self-supervised visual representation learning. Non-intrusion detection is not suitable for intrusion detection, such as \cite{65} directly using Tri-training\cite{26} in network traffic anomaly detection. So our approach takes advantage of semi-supervised learning, which requires little labeled data and has relatively high accuracy.

\vspace{-4mm}
\subsection{\textbf{Robustness of Machine Learning-based Intrusion Detection Models}}
We discuss four factors that affect the robustness of intrusion detection models (IDS), respectively.

\textbf{1) Learning from imbalanced data.} SGM \cite{73} uses an undersampling approach to re-balance. It combines the process of imbalance classes with convolutional neural networks. \textbf{2) Adversarial Attacks.} Adversarial attack methods that can preserve malicious functions are proposed in \cite{70}. The frequency domain features are used to resist adversarial attacks in \cite{56}. The impact of adversarial attacks on the model is discussed in \cite{62}, but there is no corresponding defense method. \textbf{3) Poisoning Attacks.} Pruning-based approaches \cite{42} showed that visualization approaches can aid in identifying a backdoor and poisoning attacks. It also designed pruning-based approaches to remove backdoors for Decision Trees and Random Forests. \textbf{4) Effect of data selection for training.} LDA \cite{72} focused on long-tailed distributions. It constructed the model with the long-tailed distribution as an unbalanced domain and the general distribution as a balanced domain, respectively. It then adapted models trained on the long-tailed distribution to general distributions in an interpretable way.

Studies on the first three aspects generally focus on coarse-grained attack detection based on the above findings. However, these factors are more common in fine-grained attack detection and have a heavier impact on learning the models. The fourth aspect, selecting suitable data, plays a crucial role in generalizing unseen fine-grained attacks and hence is a factor we need to consider. In this paper, we design a semi-supervised fine-grained attack classification method that can tackle the problem of imbalanced data and missing categories. Our method is independent of input features and actively selects representative samples for annotation when handling new attacks.

\vspace{-4mm}
\section{Proposed Model}



We propose a novel Weight-Task Consistency model, a semi-supervised approach to achieve label augmentation of unseen classes and high-precision multi-classification of cyber attacks on datasets with highly unbalanced data in under-labeled categories.

\vspace{-4mm}
\subsection{Overview}
\begin{figure*}[!ht]
    \setlength{\belowcaptionskip}{0pt}
    \centering
	\centerline{\includegraphics[width=0.95\textwidth]{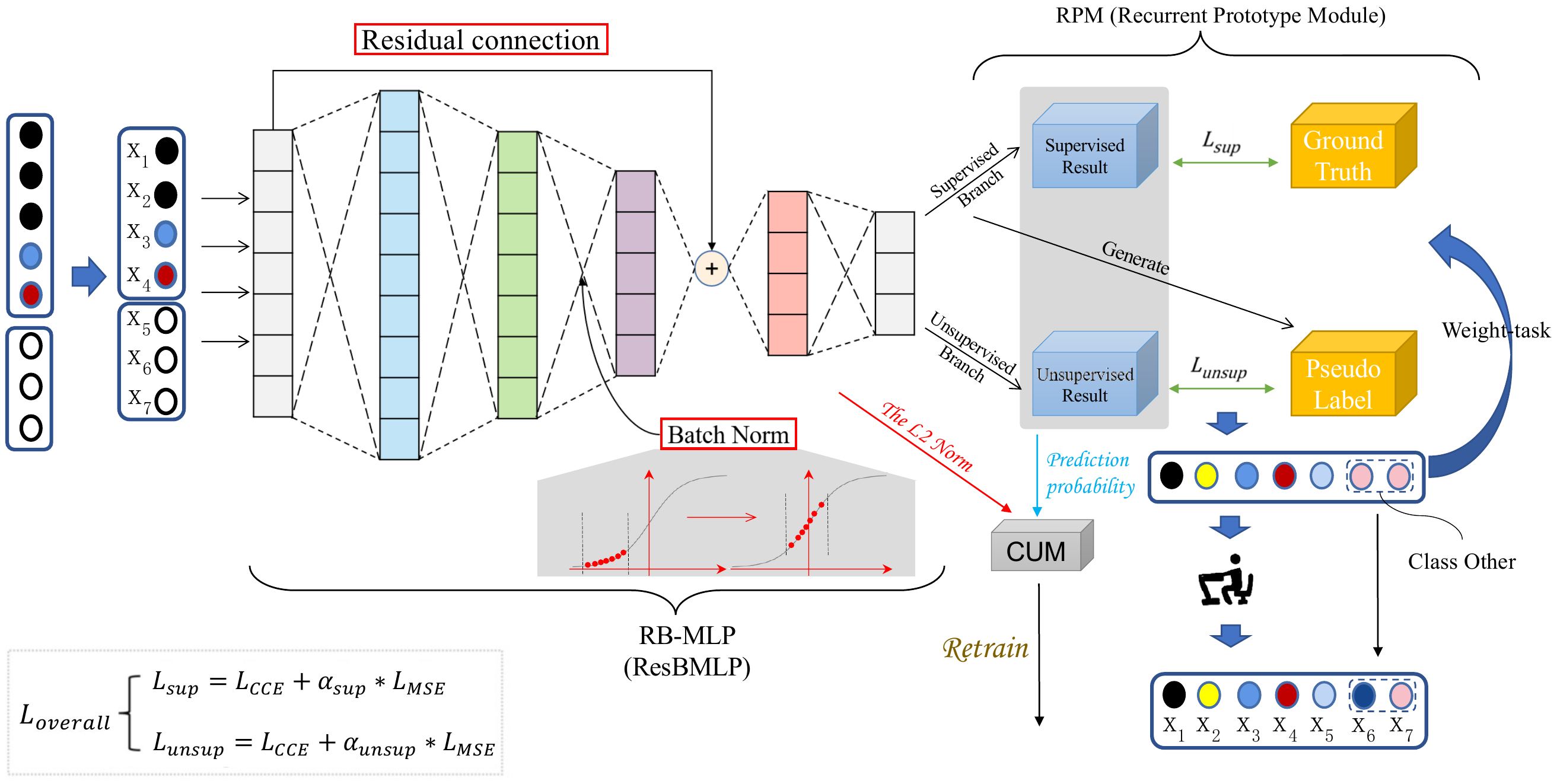}}
	\caption{Overview of the Framework (Semi-WTC). It consists of three parts, RB-MLP module, RPM module and CUM module.}
	\label{f2}
	\vspace{-5mm}
\end{figure*}

In this section, we present our proposed practical semi-supervised framework for fine-grained attack categorization (Semi-WTC) based on \textbf{W}eight-\textbf{T}ask \textbf{C}onsistency (WTC). Fig. \ref{f2} shows the overall framework of Semi-WTC, which consists of two output headers, one for the supervised branch and the other one for the unsupervised branch.

As an example, we show the classification process of our model for nine samples with low-quality labels in Fig. \ref{f2}. The first five labeled data are pre-processed and fed into the network together with the remaining three unlabeled data, where $x_{i}$ denotes the $i$-th input sample. We use different colors to distinguish different attack classes to the \emph{given labels}, where ``white'' dots $\left\{x_{5},x_{6},x_{7}\right\}$ represent samples of the attack class ``\emph{unseen}'' and ``black'' dots $\left\{x_{1},x_{2}\right\}$ represent samples of the attack class ``\emph{normal}''. Other colors denote different fine-grained attack classes, such as ``\emph{neptune}'', ``\emph{ipsweep}'', ``\emph{smurf}''. The data is then sent to the Semi-WTC model. Furthermore, in the result of the classification, the class of $x_{2}$ is changed because of the correction, $x_{5}$ has been placed in a new category, $\left\{x_{6},x_{7}\right\}$ are manually marked in the class ``\emph{other}'' to distinguish it from the other classes. That is the overall functionality of the model, as shown in Fig. \ref{f2}.

%


The overall model structure consists of the RB-MLP encoding network and the RPM module. Before being input into the network, the labeled data must be pre-processed, including random downsampling, merging few-sample labels, and regularization. After that, the pre-processed labeled data are combined with the unlabeled data as input, and the RB-MLP network encodes the semantic information of each sample. The encoding network of RB-MLP consists of two output heads, one for the supervised part and the other for the unsupervised part. Notably, the output head structures of both parts are the same to allow the parameters to be passed and shared between the supervised and unsupervised parts. Then, iterative classification results between the two branches of the two-branch structure are sequentially output through the supervised and unsupervised headers. Since the classification can be performed independently by both heads, the results of the two parts should be consistent. Therefore, we impose a Weight-Task Consistency constraint on their classification results.

\begin{table}[!ht]
\vspace{-2mm}
    \setlength{\abovecaptionskip}{0pt}
    \setlength{\belowcaptionskip}{0pt}
	\LARGE
	\caption{The list of notation in our proposed model formulation}
	\label{tab00}
	\tabcolsep 9pt 
	\renewcommand{\arraystretch}{1.2}
	\resizebox{\linewidth}{!}
	{
	\begin{tabular*}{\textwidth}{cc}
    \toprule
    Notation & Meaning \\ \hline
    $\emph{\textbf{x}}$   & Labeled and unlabeled data after pre-processing        \\
    ${\mathbf{Y}^{(m)}}$   & Output vector of the $m$-th layer in RB-MLP\\
    $Y_{i}^{(m)}$   & Output of the $i$-th node $Y_{i}^{(m)}$ in the output vector        \\
    $\mathbf{W}^{(m)}$   & Coefficient vector of the $m$-th layer in RB-MLP        \\
    \emph{BN}   & Batch Normalization        \\
    $F(.,\delta^{w}_{k})$   & Trained RB-MLP model in the $k$-th epoch        \\
    \bottomrule
    \end{tabular*}
	}
	\vspace{-2mm}
\end{table}

In addition, to fully exploit the potential information of unlabeled data, we propose the Recurrent Prototype Module (RPM). RPM aims to use the labeled information to form the primary model, generate pseudo-labels as supervised signals for unlabeled data, and conduct iterative training to optimize the encoding ability of RB-MLP for sample features continuously. Combining the re-weighting and consistency methods, we propose a weight task consistency strategy that suppresses the long-tail effect by assigning different weighting coefficients to different classes. The weight coefficients are integrated into the multiclass cross-entropy loss function, and the mean-squared error coefficient is set to combine the mean-squared difference loss function with the multiclass cross-entropy loss function. Moreover, these two loss functions are combined to form the overall loss function that supervises the whole network. In the following two subsections, we first introduce RB-MLP and the specific components of its network structure and then introduce RPM and its iterative steps.

\begin{figure}[!ht]
  \label{algorithm1}
  \renewcommand{\algorithmicrequire}{\textbf{Input:}}
  \renewcommand{\algorithmicensure}{\textbf{Output:}}
  \removelatexerror
  \begin{algorithm}[H]
    \caption{Weight-Task Consistency Algorithm}
    \begin{algorithmic}[1]
      \REQUIRE $\emph{\textbf{x}}$ $\leftarrow$ Combination of labeled and unlabeled data after pre-processing steps;
      \ENSURE The trained model of the final epoch;
      \\
      \noindent
      \textbf{Part \uppercase\expandafter{\romannumeral1}:} RB-MLP\\
      \REQUIRE $\emph{\textbf{x}}$;
      \ENSURE The $\emph{d}$-dimensional vector $Y$ representing the sample features;
      \FOR{$m=1$ to $5$}
      \STATE{$Y_{i}^{(m)}=F^{m}\left({\sum_{j=1}^N{W_{i, j}^{(m)} \times x_{i}^{(m)}}}\right);$\\
         ${\mathbf{Y}^{(m+1)}}={{F}^{m+1}}({{F}^{m}}({\mathbf{W}^{(m)}}\times {\mathbf{x}^{(m)}})+\mathbf{x});$\\
         \IF{$m=3$}
         \STATE{$Y_3=({{F}^{3}}(BN({{F}^{1\sim 2}}(x)))+x)$}
      \ENDIF
         }
      \ENDFOR
      \\
      \textbf{Part \uppercase\expandafter{\romannumeral2}:} RPM\\
      \REQUIRE ${F(.,\delta^{w}_{k=0})}$;
      \ENSURE Updated weights of RB-MLP $\gets$ Trained RB-MLP model  
      \WHILE {\emph{NOT} meet the conditions for early stop}
      \STATE Train a prototype RB-MLP model ${F(.,\delta^{w}_{k=0})}$ (if prediction probability $\leq$ 75\%, then CUM) $\gets$ labeled data;
      \STATE Generate pseudo labels using $F(.,\delta^{w}_{k=0})$;
      \STATE Train a RB-MLP model $F(.,\delta^{w}_{k=k+1})$ $\gets$ unlabeled data (pseudo labels);
      \ENDWHILE
      \\
    \textbf{return} Final epoch with high classification accuracy;	
    \end{algorithmic}
  \end{algorithm}
\end{figure}
\vspace{-2 mm}

\vspace{-3mm}
\subsection{Residual Connection \& Batch Norm Multilayer Perceptron (RB-MLP)}
Since traditional machine learning methods in intrusion detection naturally target learning the distribution of a few categories with difficulties, there are more misclassifications. In addition, traditional misuse detection methods also have an inherent flaw in that they classify intrusion categories by matching intrusion behaviours with existing intrusion patterns. This approach is powerless against unseen intrusion attacks. In summary, we decide to use a neural network to construct the feature encoder, and our RB-MLP encoder structure is shown in Fig. \ref{f2}, which mainly consists of five fully connected layers. Moreover, we introduce residual connectivity and Batch Normalization to it. The steps of the entire RB-MLP are as follows. First, the unbalanced training set is divided into training and validation sets by random down-sampling
of samples while setting the test set. The data in the training set are then fed into the RB-MLP model to generate a set of $N$ $d$-dimensional output embeddings, which are then proposed as $d$-dimensional vectors to represent the sample features and finally provided to the two-branch classifier in RPM to predict the labels of the original intrusions.

The neural network has the advantages of high detection accuracy with excellent nonlinear mapping and self-learning capability, simple modeling, and other advantages. At the same time, the neural network has a strong ability to analyze attack patterns, which is suitable for intrusion detection requirements in terms of concept and processing methods. It has become one of the hot spots for research in the field of intrusion detection. Convolutional neural networks have made significant progress in trend prediction and target detection. However, since network intrusions are often recorded as traffic, their data samples may lack contextual associations and thus are unsuitable for feature extraction using $one$-dimensional convolution. In this regard, we use an artificial neural network, i.e., MLP (Multilayer Perceptron) model, as the feature encoding network. There are multiple hidden layers inside the MLP itself, and the layers are fully connected. In this way, semantic feature information between contexts can be transferred. This structure is beneficial for intrusion traffic, which has multidimensional features, and information of different dimensions is often challenging to transfer effectively. In our proposed RB-MLP, through comparison experiments, the number of hidden layers is finally determined as five, with dimensions of 256, 128, 64, 32, and the total number of predicted categories, respectively. Except for the Softplus activation function used after the first fully connected layer, we perform a ReLU activation after each of the remaining fully connected layers. 
\vspace{-0.3mm}
\begin{equation}
\begin{aligned}
	&Y_{i}^{(m)}=F^{m}\left({W^{(m)}_{i, 1} \times x^{(m)}_{1}+W^{(m)}_{i, 2} \times x^{(m)}_{2}+\cdots}\right. \\
	&\left.{+W^{(m)}_{i, n} \times x^{(m)}_{n}+\cdots}\right)
	\label{eq1}
\end{aligned}
\end{equation}

The expression of the $m$-th fully connected layer is shown in Eq. (\ref{eq1}), where $x_{n}^{(m)}$ represents the $n$-th feature of the $m$-th fully connected layer and $W_{i, n}^{(m)}$ represents the coefficient corresponding to the feature, so we can calculate the output of the $i$-th node $Y_{i}^{(m)}$ in the output vector ${\mathbf{Y}^{(m)}=\big(Y_{1}^{(m)}, Y_{2}^{(m)}, \ldots, Y_{i}^{(m)}, \dots\big)}^\mathrm{T}$ from Eq. (\ref{eq1}).
\vspace{-0.3mm}
\begin{equation}
	{\mathbf{Y}^{(m+1)}}={{F}^{m+1}}({{F}^{m}}({\mathbf{W}^{(m)}}\times {\mathbf{x}^{(m)}})+\mathbf{x})
	\label{eq2}
\end{equation}
\vspace{-0.3mm}
Inspired by the recently proposed Deep Crossing \cite{37},
which has achieved good results in feature extraction by introducing
residual units between embedding layers, we also add residual structure to the MLP, as shown in Eq. (\ref{eq2}). The $\mathbf{x}^{(m)}$ is the original feature of the sample ${\mathbf{x}^{(m)}=\big(x_{1}^{(m)}, x_{2}^{(m)}, \ldots, x_{n}^{(m)}, \ldots\big)}^\mathrm{T}$ and the corresponding parameter is $\mathbf{W}_{i}^{(m)}=\big(W_{i, 1}^{(m)}, W_{i, 2}^{(m)}, \ldots, W_{i, n}^{(m)}, \ldots\big)$. So the main learnable parameter of the $\emph{m}$-th layer is ${\mathbf{W}^{(m)}=\big(\mathbf{W}_{1}^{(m)}, \mathbf{W}_{2}^{(m)}, \ldots, \mathbf{W}_{n}^{(m)}, \ldots\big)}^\mathrm{T}$. We set the number of layers to 3, i.e. $m=3$. $W_{i}$ mixes all feature information from this traffic sample. We use the residual structure to combine the original feature information with the encoded features, effectively mitigating the loss of feature semantic information. In addition, due to the significant difference between feature extremes, Batch Normalization is introduced between the second and third fully-connected layers for better encoding of features and combining the original features of the samples. The whole RB-MLP network expression is shown in Eq. (\ref{eq3}), where $BN$ stands for Batch Normalization.

\begin{equation}
\setlength{\abovedisplayskip}{1pt}
\setlength{\belowdisplayskip}{1pt}
	Y={{F}^{4\sim 5}}({{F}^{3}}(BN({{F}^{1\sim 2}}(x)))+x)
	\label{eq3}
\end{equation}

\vspace{-4mm}
\subsection{Recurrent Prototype Module (RPM)}
\begin{figure*}[!ht]
    \setlength{\belowcaptionskip}{0pt}
	\centerline{\includegraphics[width=0.90\textwidth]{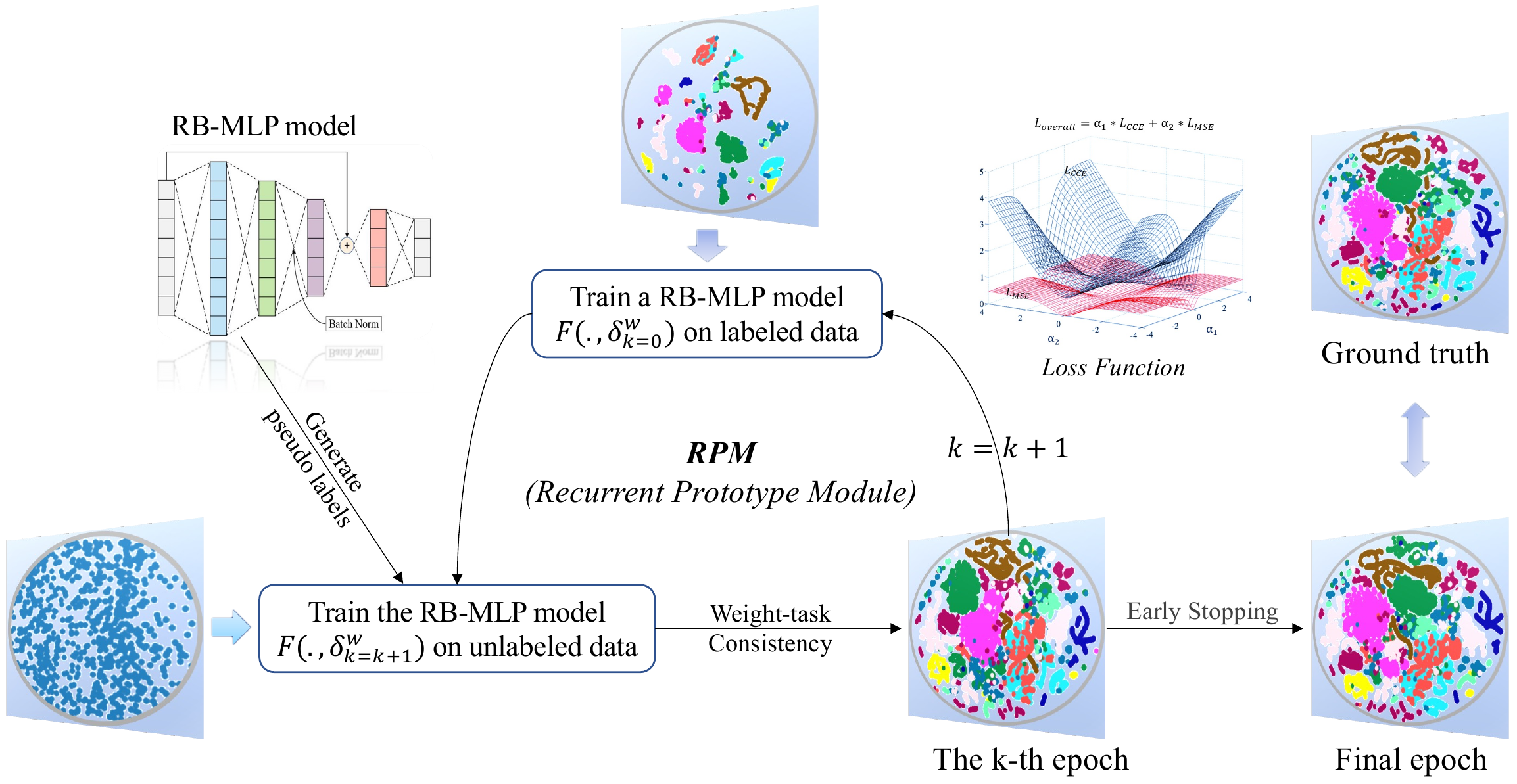}}
	\caption{Overview of the Recurrent Prototype Module (RPM). It consists of generating pseudo labels, using pseudo-labels to train RB-MLP with unlabeled data and determining whether the early stop condition is met. Here different colors here represent different categories. RPM can make the model learn the distribution of labeled and unlabeled data simultaneously.}
	\label{f3}
	\vspace{-3 mm}
\end{figure*}

Due to the specific properties of different traffic samples, the classification results have problems such as large sample size and a high false alarm rate for intrusion detection, which is challenging for intrusion detection and classification. Moreover, existing methods are highly dependent on a large number of labeled samples in achieving high accuracy classification. Recently, semi-supervised learning has been increasingly applied in various tasks, which are mainly concerned with making good use of a large amount of unlabeled data and a small amount of labeled data for training. It can effectively alleviate the dependence of models on labeled samples. Therefore, we aim to train our RB-MLP network in a semi-supervised manner to achieve traffic intrusion sample classification, for which we propose the RPM (Recurrent Prototype Module), as shown in Fig. \ref{f3}.

Inspired by the multilevel attention-based prototype network (MapNet) \cite{38}, we similarly design a recurrent prototype module (RPM). Using RPM, we can make the model learn the distribution of labeled and unlabeled data simultaneously. In addition, to make the distributions of labels predicted by the two branches consistent, \textbf{Weight-Task Consistency} is introduced into the iterative process. This consistency is specified by setting different weight coefficients $\delta_{i}$ for different categories in the loss function according to the number of each category in each batch to suppress the effect from the long-tail distribution. The steps of the whole RPM are shown in Fig. \ref{f3}.

\textbf{Step 1:} A prototype RB-MLP is trained by labeled data.

\textbf{Step 2:} Then, the prototype RB-MLP is used to generate pseudo labels for unlabeled data.

\textbf{Step 3:} Pseudo-labels are used as supervised information to train RB-MLP with unlabeled data.

\textbf{Step 4:} Determine if the early stop condition is met. If not, repeat Steps 1-3 above.

It is worth noting that the structure of the supervised and unsupervised parts of the network is identical in RPM. The difference lies in the presence or absence of data labels, which allows the model to share parameters and fit both labeled and unlabeled data. For the design of the loss function, we design a combined loss function as shown in Eq. (\ref{eq5}). 
\vspace{-0.3mm}
\begin{equation}
	{L}_{overall}={{L}_{CCE}}+\alpha \times {L}_{MSE}
	\label{eq5}
\end{equation}
\vspace{-0.3mm}

The loss function consists of two components, the mean squared error loss function ${{L}_{MSE}}$ and the multiclass cross-entropy loss function ${{L}_{CCE}}$, where $\alpha$ represents the weight of the mean squared error loss function ${{L}_{MSE}}$, as shown in Eq. (\ref{eq6}). These two loss functions focus on different directions. Moreover, they can complement each other to a large extent \cite{yeung2021mixed}, especially in the case of semi-supervised learning, existing labels can produce bias on the model, and the two loss functions can avoid the phenomenon.

\vspace{-0.3mm}
\begin{equation}
	{{L}_{MSE}}=\frac{1}{N}{{\sum\limits_{i}{({{y}_{i}}-{{p}_{i}})}}^{2}} 
	\label{eq6}
\end{equation}
\vspace{-0.3mm}

For the supervised part $\alpha=\alpha_{sup}$ and the unsupervised part $\alpha=\alpha_{unsup}$, ${{L}_{CCE}}$ represents the multiclass cross-entropy loss function, as shown in Eq. (\ref{eq7}). And $N$ represents the number of samples and ${{y}_{i,j}}$ is the sign function, which takes the value of $1$ if sample $i$ is in the correct category $j$. Otherwise, the value is 0. ${p}_{i,j}$ is equal to the probability that the model predicts that sample $i$ belongs to category $j$.

\begin{equation} 
    {{L}_{CCE}}=\frac{1}{N}\sum\limits_{i}{{{L}_{CCE,i}}=-}\frac{1}{N}\sum\limits_{i}{\sum\limits_{j=1}^{C}{{{y}_{i,j}}\log ({{p}_{i,j}})}}
	\label{eq7}
\end{equation}
\begin{equation}
	{{L}_{WTC}}=-\frac{1}{N}\sum\limits_{i}{\sum\limits_{j=1}^{C}{{{y}_{i,j}}{{\delta }_{j}}\log ({{p}_{i,j}})}}+ \frac{\alpha}{N}{{\sum\limits_{i}{({{y}_{i}}-{{p}_{i}})}}^{2}}
	\label{eq8}
\end{equation}

Combining the Weight-Task Consistency and the loss function ${L}_{overall}$, we design the WTC loss function ${{L}_{WTC}}$, as shown in Eq. (\ref{eq8}). And ${\delta}_{j}$ denotes different weight coefficients set for different categories in the loss function, which also reflects Weight-Task Consistency in the iterative process.

\vspace{-4mm}

\subsection{Class Uncertainty Module (CUM)}
\begin{figure}[!ht]
\vspace{-3mm}
    \setlength{\abovecaptionskip}{0pt}
    \centerline{\includegraphics[width=\columnwidth]{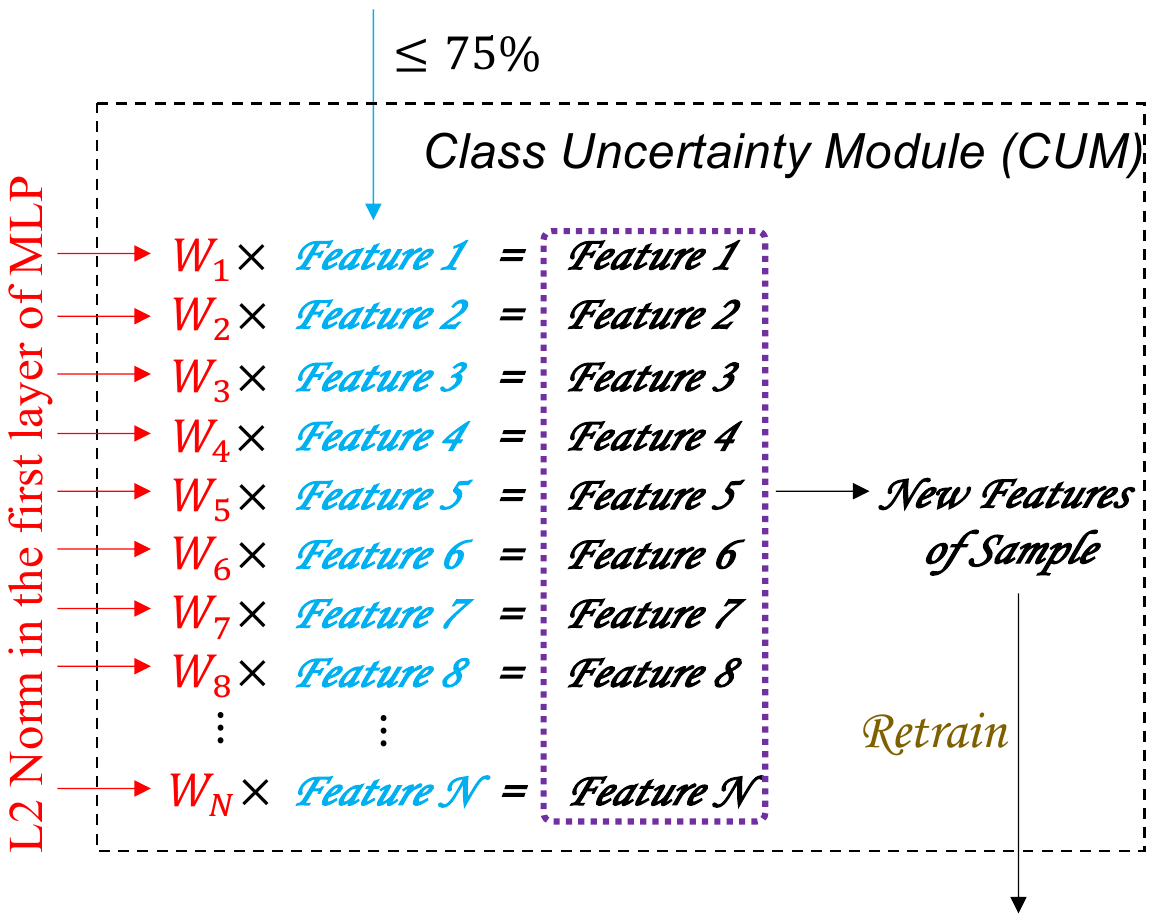}}
	\caption{Overview of the Class Uncertainty Module (CUM). The weight coefficient $W_{i}$ of each feature is multiplied with the feature $i$, thus enhancing the important features, suppressing the influence of irrelevant features, and reducing the uncertainty samples at the category classification boundary.}
	\label{f4-1}
	\vspace{-3mm}
\end{figure}

In the long-tail problem, uncertainty is considered the essential factor affecting the model performance. As for the attack categorization, we believe that the various sample features will cause the uncertainty of the model. And this is because uncertain samples are often at the classification boundary of different categories, so it is necessary to strengthen important features and suppress irrelevant features during the training phase. Therefore, we propose the Class Uncertainty Module (CUM) to address the above issues, as shown in Fig. \ref{f4-1}.

We first judge whether the sample belongs to the uncertain sample by the prediction probability of the category (the prediction probability is no more than 75\%). If the prediction probability is no more than 75\%, the sample will be included in the CUM for processing. Otherwise, it will not be processed. The processing method for labeled and unlabeled data is the same, in which the critical features of each sample need to be judged to change the original feature input. The weight coefficient of each feature is the L2 norm between the input layer and the first layer of MLP, and then the coefficient is multiplied by the feature to suppress the influence of irrelevant features.

\vspace{-4mm}
\subsection{Rationalization of Model Structure}
\subsubsection{\textbf{Reasonableness of Feature-independence}}
Three common feature extraction methods are as follows: \cite{56} \cite{57} are based on flow-based features, \cite{58} are based on packet-based features, while \cite{60} \cite{61} are based on byte-code based feature extraction. Our method can be seen as secondary feature extraction based on previous feature extractions, and features can be extracted again after they have been extracted, so what kind of features they are based on can be regardless.

\subsubsection{\textbf{Reasonableness of RB-MLP}}

\textbf{(1)} For different feature extraction methods, such as secondary features encoded by various auto-encoders, there is almost no difference in performance between the original features and the encoded secondary features. The reason is that our RB-MLP structure can function as an auto-encoder with its first three layers of the network. As a result, this can improve the model's ability to encode local features.

\textbf{(2)} RB-MLP fuses the high-level and low-level information, so the extracted features are more comprehensive to achieve high-precision classification.

\subsubsection{\textbf{Reasonableness of RPM}}
\textbf{(1)} The introduction of the RPM loop structure allows for iterative learning of data label information and full exploitation of potential information in unlabeled data, which provides better robustness for poisoning attacks. 

\textbf{(2)} The introduction of the category weight coefficient $\delta_{i}$ enables the Semi-WTC model to effectively control the influence of significant classes and thus increase the ability to identify minor classes.

\textbf{(3)} RPM can effectively perform feature extraction for unlabeled data and gradually fit the unseen samples to their accurate labels through iterative operations by introducing pseudo-labeled supervisory signals.

\subsubsection{\textbf{Reasonableness of CUM}}
\textbf{(1)} The CUM can alleviate the problem of model uncertainty in the long-tail distribution by using the feature re-weighting method.

\textbf{(2)} Since the prediction probability is used as an indicator to measure the uncertainty of the sample, the calculation cost will not be increased. Therefore, Semi-WTC remains a cost-friendly model.

\subsubsection{\textbf{Reasonableness of Our Semi-WTC Model}}
\textbf{(1)} Our Semi-WTC chooses fully connected layers plus residual connections as the basis for network construction because of its good fault tolerance and nonlinear global action, which can better represent the distribution of unseen data. Our Semi-WTC is a general semi-supervised framework that can combine with any neural network structure.

\textbf{(2)} Our Semi-WTC model introduces a two-branch structure to extract the semantic information of features for both labeled and unlabeled data, so it can perform better than the IDS-MLP, which belongs to the same perceptron model.

\textbf{(3)} The Semi-WTC has high execution efficiency in large datasets and can efficiently identify different traffic features. Compared to clustering, this excellent performance is due to the smaller parameters of the multilayer perceptron and the shallow network layers of the Semi-WTC itself.

\textbf{(4)} Our Semi-WTC introduces a category weight through the weight consistency function, which can effectively balance the long-tail classes and thus improve the performances of fine-grained attack detection on tail classes.

\subsubsection{\textbf{Reasons for the Improvement in Robustness of Our Semi-WTC Model}}
\textbf{(1)} Semi-WTC trains the initial model by labeling information. Moreover, fusing the information of the supervised and unsupervised parts separately through a two-branch structure enhances the model's robustness. Even if some labels in the supervised part are mislabeled, the pseudo-labels in the unsupervised part can help correct the model and explore the potential information inside the data \cite{21}.

\textbf{(2)} The introduction of the RPM loop structure in our Semi-WTC model can learn the potential information inside the data, which has better robustness for the poisoning attack. Also, under adversarial attacks, our Semi-WTC has good performance because of its two-branch structure, which can extract semantic information of features for both labeled and unlabeled data.
\section{Active Adaption Resampling}
\begin{figure}[!ht]
    \vspace{-2mm}
    \setlength{\belowcaptionskip}{0pt}
	\centerline{\includegraphics[width=\columnwidth]{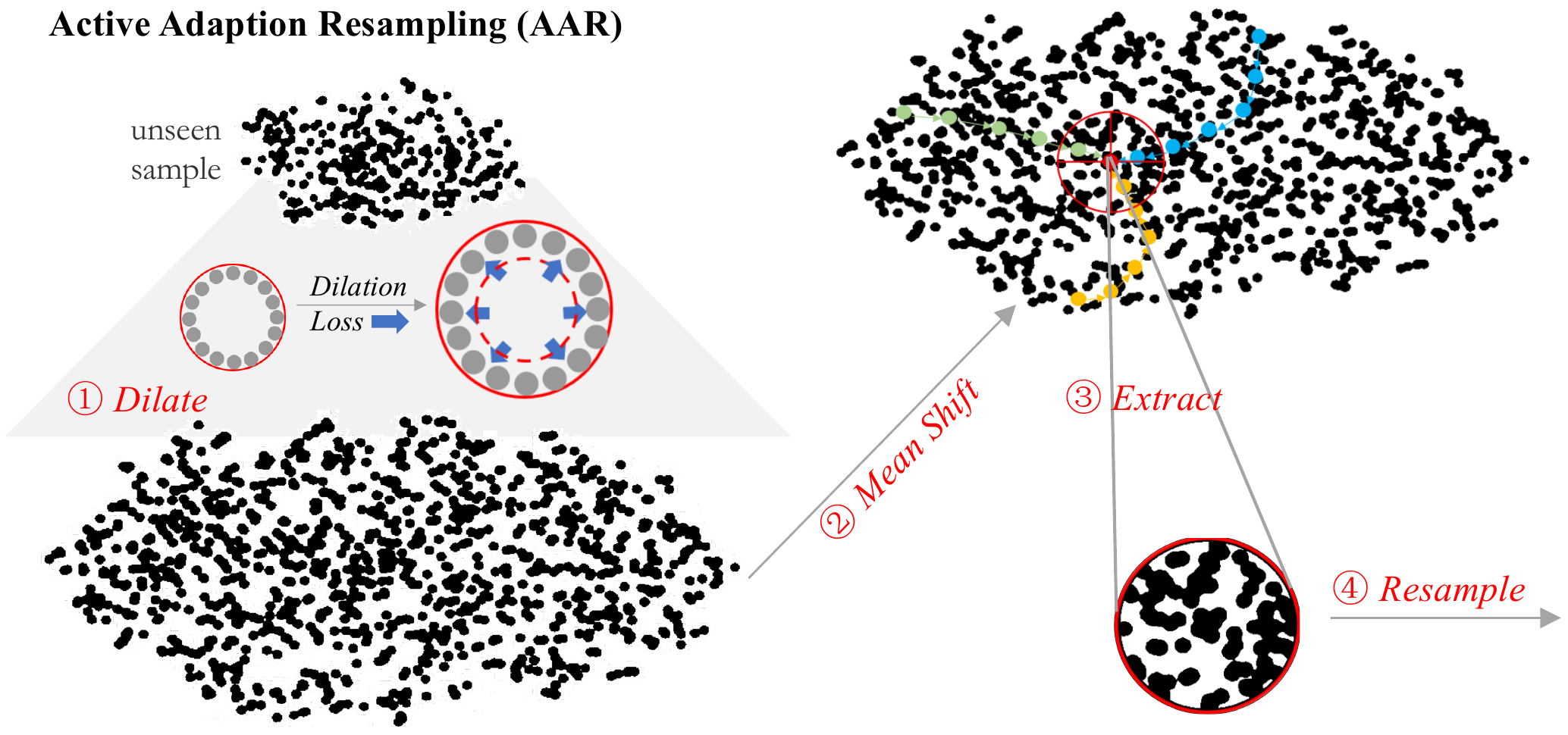}}
	\caption{Specific steps of active adaption resampling method}
	\label{f4}
	\vspace{-4mm}
\end{figure}

New traffic attacks are emerging in the real world, and most of these brand-new attacks cannot be detected by existing intrusion detection systems since their feature distribution is often inconsistent with previous anomalous traffic. For this reason, people need to label these unseen samples, but how to better select a limited number of valid samples for labeling becomes a challenge that cannot ignore. Since our motivation is to detect anomalous traffic attacks and reduce the false alarm rate as much as possible, we resample the core samples that characterize the class distribution rather than the classification decision boundaries usually chosen by traditional active learning methods. In addition, we design a Dilation loss function to speed up finding the dense center of Mean Shift by expanding the distribution of samples. We can then discover the distribution of feature samples around the dense center. To this end, we develop the Active Adaption Resampling (AAR) method, as shown in Fig. \ref{f4}. Next, we will describe the AAR in detail.

\begin{table*}[!ht]
    \setlength{\abovecaptionskip}{0pt}
    \setlength{\belowcaptionskip}{0pt}
    \centering
	\small
	\caption{The specific division of the Datasets}
	\label{tab1}
	\tabcolsep 12pt 
	\renewcommand{\arraystretch}{1.2}
	{
	\begin{tabular*}{\textwidth}{ccccc}
	\toprule
	Division of Dataset	& NSL-KDD\cite{9} & CIC-IDS 2018\cite{10} & CIC-IDS 2018 (big)\cite{10} & AndMal 2020\cite{27}\cite{28} \\\hline
	  Train set(labeled) & 223& 192& 6696 & 119 \\
	  Train set(unlabeled) & 22077& 18994 & 662934 & 11755 \\
	  Validation set & 5575& 4797 & 167408 & 2969 \\
	  Test set & 4897& 5995 & 209260 & 3711 \\
	  Total  & 32772& 29978 & 1046298 & 18565   \\
	\bottomrule
	\end{tabular*}
	}
	\vspace{-4 mm}
\end{table*}
Specifically, AAR can be divided into four stages: Dilation, Mean Shift\cite{47}, Extraction, and Resampling.

\textbf{1. Dilation:} Using a linear layer, we change the data distribution of the unseen samples to make different vectors more divergent. We refer to the work of Peng et al. \cite{45} to construct our Dilation Loss using Hinge Embedding Loss, which can help determine whether two feature vectors are similar. Unlike \cite{45}, we also make use of \emph{t-SNE} so that a visual representation of the sample features can be extracted, as shown in Fig. \ref{f4}. The inter-class distance between class $i$ and $j$ is defined as $d_{inter}=1-\cos \left(X_{i}, X_{j}\right)$, where $X_{i}$ and $X_{j}$ are the feature vectors of categories $i$ and $j$. By defining $N$ as the number of classes that appear in a mini-batch, we need to minimize the Dilation Loss $L_{dilation}$ aims to maximize $d_{inter}$. It is a variant of hinge loss with $MARGIN$, which is 1 in our AAR unless otherwise noted. The expression of Dilation Loss $L_{dilation}$ is shown as Eq. (\ref{eq9}), where $\mathcal{M}$ represents $MARGIN$.
\vspace{-0.3mm}
\begin{equation}
	L_{\text {dilation}}=\sum_{i=1}^{N} \sum_{j \neq i, j=1}^{N} \max \left\{ 0, \mathcal{M}-\left(1-\cos \left(X_{i}, X_{j}\right)\right)\right\}
	\label{eq9}
\end{equation}

\textbf{2. Mean Shift:} Mean shift algorithm \cite{47} is a stepwise optimization algorithm based on kernel density estimation. Given a set $X$ of $n$ data points in a $d$-dimensional space, the primary form of the mean shift vector for any point $x$ in the space can be expressed as Eq. (\ref{eq10}):

\begin{equation}
\setlength{\abovedisplayskip}{1pt}
\setlength{\belowdisplayskip}{1pt}
	M_{h}(x)=\frac{1}{k} \sum_{x_{i} \in S_{h}}\left(x_{i}-x\right)
	\label{eq10}
\end{equation}
\noindent
where $x_{i}$ is $n$ sample points, $i=1,2,\cdots,n$, $S_{h}(x)=\left\{y:\left(y-x_{i}\right)^\mathrm{T}\left(y-x_{i}\right)<h^{2}\right\}$ is a high-dimensional sphere of radius $h$ centered at $x$, denoting the effective region, which contains $k$ sample points. Eq. (\ref{eq10}) can also be expressed as Eq. (\ref{eq11}):
\begin{equation}
	\hat{x}=x+M_{h}(x)=\frac{1}{k} \sum_{x_{i} \in S_{h}} x_{i}
	\label{eq11}
\end{equation}
\noindent
Based on above equations, $M_{h}(x)$, as the offset mean vector of $x$, can be used to update $x$ for $\hat{x}$. After adding the kernel function, Mean Shift clustering means that the element is marked as the same class at its convergence position for each element in the set. The process keeps repeating until convergence. Based on the visual representation of \emph{t-SNE} obtained, the Mean shift can be used to find the dense center of the batch of sample data. We believe that the samples near the dense center are the core samples that can reflect the characteristics of the category distribution.

\textbf{3. Extraction:} According to the dense center found in the previous stage, we draw several samples around it as few of the number of current unseen samples, and we can know that the batch of samples can approximate the data distribution of the current unseen samples. Here we treat the unknown class as a whole, and if there are multiple classes, we estimate the distribution of the most dominant class of the unknown classes.

\textbf{4. Resampling:} After that, the extracted samples are re-labeled and fed into the labeled training set, and then the whole model is trained to update the parameters of the RB-MLP network.

\vspace{-4mm}
\section{Experiments and Evaluation}
\subsection{Metrics for Evaluating Semi-WTC}
For the evaluation metrics, accuracy and Macro-F1 are adopted to evaluate the performance of our model, which can reflect the performance of the classification model more comprehensively. From the confusion matrix, true positive ($TP$), true negative ($TN$), false positive ($FP$), and false negative ($FN$) values are obtained. Then accuracy, as well as Macro-F1, are calculated as Eq. (\ref{eq12}) and Eq. (\ref{eq14}). The evaluation metrics are more complex for network traffic anomaly detection because the data sets are primarily unbalanced with more regular and less abnormal data. Here we also use TPR (True Positive Rate) metrics to evaluate the performance and performance of models and so on, which can consider the number of categories and the distribution of datasets, calculated as in Eq. (\ref{eq-add1}). In Eq. (\ref{eq14}) and Eq. (\ref{eq-add1}), $TP_{i}$ represents the true positive value of class $i$, and same for $FP_{i}$ and $FN_{i}$.

\begin{equation}
\setlength{\abovedisplayskip}{1pt}
\setlength{\belowdisplayskip}{1pt}
	Accuracy=(TN+TP)/(TN+FN+TP+FP)
	\label{eq12}
\end{equation}

\begin{equation}
\setlength{\abovedisplayskip}{1pt}
\setlength{\belowdisplayskip}{1pt}
	Macro-F1=\frac{\sum\nolimits_{i=1}^{C}{2T{{P}_{i}}/(2T{{P}_{i}}+F{{P}_{i}}+F{{N}_{i}})}}{C}
	\label{eq14}
\end{equation}

\begin{equation}
\setlength{\abovedisplayskip}{1pt}
\setlength{\belowdisplayskip}{1pt}
	TPR=\frac{\sum\nolimits_{i=1}^{C}{T{{P}_{i}}/(T{P}_{i}+F{N}_{i})}}{C}
	\label{eq-add1}
\end{equation}
\vspace{-8mm}

\subsection{Datasets and Experimental Environment}
\subsubsection{\textbf{Datasets}}
We evaluate the proposed approach on the following datasets: NSL-KDD Dataset, CIC-IDS 2018 Dataset, and AndMal 2020 Dataset.

\textbf{NSL-KDD Dataset}\cite{9}: The KDD Cup 99 dataset\cite{hettich1999kdd} has been widely used for intrusion detection. It is a stream-based dataset and consists of approximately 4,900,000 sample vectors, each containing 41 features and four classes of attacks: Denial of Service (DoS), Probe, Remote to Local (R2L), and User to Root (U2R). Tavallaee et al. proposed the NSL-KDD dataset \cite{9} based on the KDD Cup 99 dataset \cite{hettich1999kdd}. NSL-KDD dataset removes a large amount of redundant data from the KDD Cup 99 dataset, which can greatly ensure data validity. There are no duplicate records in its test set, so redundant records will not affect the classifier, and computational cost is low.

\textbf{CIC-IDS 2018 Dataset}\cite{10}: The CSE-CIC-IDS 2018 dataset consists of records that simulate normal traffic in a small network environment. On the other hand, CIC-IDS 2018 dataset \cite{10} is prepared on a more extensive network. This dataset includes seven attack scenarios: Brute Force Method, Heartbleed, Botnet, DoS, DDoS, and Web Attack. These attacks penetrate the network from within. The dataset also includes some features covering a portion of the evaluation field. We utilize this dataset in our experiments since it is representative of modern network traffic.

\textbf{AndMal 2020 Dataset}\cite{27}\cite{28}: UNB presents a new large android malware dataset for research in detecting android malware in smartphones called CIC-AndMal-2020. The dataset includes 200,000 benign samples and 200,000 malware samples for 400,000 Android applications, with 14 prominent malware categories. There are 15 categories in the AndMal 2020 dataset, including 14 types of malware and one category of normal software. Each software sample has 9441-dimensional features such as Intents Actions, which we can treat similarly to sample features such as IP addresses on the traffic dataset.

\vspace{-4mm}
\subsubsection{\textbf{Specific Experimental Environment}}
Our proposed approach is implemented by the deep learning framework Keras, PyTorch, and related Python libraries, such as Sklearn and NumPy. Our experiments are conducted on the Kaggle platform and the server. The main parameters of the server are as follows: the operating system is Ubuntu 16.04.4 LTS, the CPU is Intel(R) Xeon(R) CPU E5-2665 0 @ 2.40GHz, the GPU is GTX 1080Ti, and the memory capacity is 96 GB. 
\begin{table*}[!ht]
    \setlength{\abovecaptionskip}{0pt}
    \setlength{\belowcaptionskip}{0pt}
	\footnotesize
	\caption{Experimental results on the NSL-KDD dataset with 1\% labeled training data}
	\label{tab3}
	\tabcolsep 1.75pt 
	\renewcommand{\arraystretch}{1.2}
	\resizebox{\linewidth}{!}
	{
	\begin{tabular*}{\textwidth}{ccccccccccccccc}
	\toprule
	  Model&Acc (\%)& Macro-F1 (\%) & TPR (\%)&back&ipsweep&neptune&nmap&normal&other&portsweep
	  &satan&smurf&teardrop&warezclient \\\hline
	  KMeans\cite{29} &59.83	&55.74 &57.91	&80.6	&81.2	&88.7	&40.9	&88.9	&38.9	&81.3	&79.2	&90.6	&82.6	&45.9\\
	  RF\cite{30} &85.54	&81.06 &83.47	&97.4	&61.3	&98.8	&69.8	&92.4	&46.4	&92.3	&95.1	&99.1	&96.3	&91.1\\
	  SVM\cite{31} & 86.23	&82.32 &84.28	&97.3	&67.3	&99.4	&65.6	&90.3	&50.6	&90.7	&94.5	&98.2	&96.9	&87.6\\
	  LR/SGD\cite{32} &82.74	&80.77 &80.57	&97.2	&58.2	&99.7	&66.3	&90.8	&44.1	&93.7	&95.7	&99.7	&97.7	&89.6\\
	  LSTM\cite{33} &84.91	&83.12 &82.80	&96.8	&74.5	&99.6	&70.5	&93.4	&45.7	&89.9	&92.5	&99.2	&97.5	&89.2\\
	  VGGNet\cite{34} &89.03	&86.92 &87.11	&97.1	&84.4	&99.1	&80.4	&92.3	&66.7	&93.5	&95.3	&97.9	&97.8	&92.3\\
	  ACID\cite{15}   &91.05	&88.91 &88.93	&97.7	&92.1	&94.9	&71.5	&91.2	&72.3	&94.6	&92.5	&97.6	&98.2&90.5\\
	  FARE-unsup\cite{12} &67.12&63.31 &65.25 &85.2	&87.7	&90.5	&42.8	&89.7	&40.5	&88.6	&84.7	&94.2	&86.4	&48.2\\
	  FARE-semisup\cite{12}  &88.21&82.57 & 86.14 &96.9	&95.0	&\textbf{99.8}	&44.7	&92.1	&45.7	&90.8	&91.4	&\textbf{100.0}	&97.8	&63.7\\
	  MixMatch\cite{44} &91.23&89.76 & 89.79&97.9	&92.7	&97.9	&74.3	&90.6	&63.3	&\textbf{96.9}	&89.8	&98.0	&\textbf{98.7}	&93.0\\
	  Semi-WTC   &\textbf{94.33}	&\textbf{92.73}  &\textbf{92.37}	&95.3	&\textbf{97.5}	&98.7	&\textbf{92.1}	&\textbf{93.3}	&\textbf{77.1}	&95.8	&\textbf{96.3}	&99.6	&98.1	&\textbf{94.5}\\
	  Support\footnote{Support means the number of samples in the training set for each category.}  &- &-  &- &773	&2891 &4024	&1185 &3973	&669 &2319 &2909 &2134	&695 &728\\
	\bottomrule
	\end{tabular*}
	}
	\vspace{-2 mm}
\end{table*}

In addition, for datasets not specifically divided into test sets, such as the CIC-IDS2018 dataset and the AndMal 2020 dataset, we randomly split 80\% of the data into the training set and 20\% into the test set. 20\% of the training set is used as the validation set, and the remaining 80\% is divided into the labeled training set and unlabeled training set according to a certain ratio. The number of samples in the specific data set is shown in Table \ref{tab1}. The ablation experiments are also performed on the ratio of labeled data. For the CIC-IDS 2018 dataset, if ``big'' is not explicitly indicated, it generally refers to the CIC-IDS 2018 dataset with small sample size. The Adam optimizer ($learning$ $rate=0.001$, $\beta_{1}=0.9$, $\beta_{2}=0.999$) is used uniformly to execute multiple epochs in the training phase until the model performance no longer improves, while different batch sizes (default is 2,000) are set for different datasets to accommodate different data sizes.

\vspace{-4mm}
\subsection{Data Pre-processing}
\label{prepro}
Due to sampling errors, some datasets contain noise samples, missing and infinity values, and other disturbing items. Therefore, pre-processing the original dataset is crucial in subsequent data mining and feature engineering. Taking the NSL-KDD dataset \cite{9} as an example:

\textbf{Step 1:} We first randomly downsample the dataset by setting a fixed random seed, especially for head classes in the long-tail distribution such as the class ``\emph{normal}'' and class ``\emph{neptune}''.

\textbf{Step 2:} Since some categories of target labels are rarely observed, and certain labels are only in the test set, we keep the labels of certain categories of attacks unchanged and change the labels to \textbf{``\emph{other}''} for all other attacks. Specifically, we will keep the ten categories of labels: ``\emph{normal}'', ``\emph{neptune}'', ``\emph{satan}'', ``\emph{ipsweep}'', ``\emph{portsweep}'', ``\emph{smurf}'', ``\emph{nmap}'', ``\emph{back}'', ``\emph{teardrop}'', and ``\emph{warezclient}''. Then, we change all the remaining labels to ``\emph{other}''.

\textbf{Step 3:} The logarithmic operation is performed for some numerical data. It can eliminate the influence of dimensionality between indicators and speed up gradient descent and model convergence. The formula for this operation is $X=\log (X+\xi )$, where $\xi$ is a minimal quantity used to ensure that ${X+\xi>0}$ so that the logarithmic operation is always valid. In addition, the results of the logarithmic process are related to the overall sample distribution, which means each sample point will influence the logarithmic operation.

\textbf{Step 4:} Afterwards, one-hot encoding is performed for the features. The NSL-KDD dataset contains three non-numeric (categorical) features: protocol type, service, and flag. These three features will yield 84 encoded features using one-hot encoding\footnote[2]{``Support'' indicates the number of samples in training set for each category.}.

\textbf{Step 5:} Finally, the processed training set is further divided into three parts, labeled training set, unlabeled training set, and validation set. The pre-processing steps are similar for the other two datasets, CIC-IDS 2018 and AndMal 2020. It is worth noting that when processing the AndMal 2020 dataset, some dynamic features, such as $Runtime\_exec$ in the dataset with a more dispersed distribution, are removed to extract similar features between samples of the same class better.

\vspace{-4mm}
\subsection{Regular Experiments for Semi-WTC}

\subsubsection{\textbf{Comparison with SOTA Methods}}
In our experiments, FARE \cite{12}, ACID \cite{15}, and traditional machine learning algorithms are chosen as baseline methods for comparison. FARE and ACID are recently proposed methods with high citations. FARE \cite{12} is a clustering method that enables intrusion attack classification with low-quality labels. We implemented two FARE types for comparison: the unsupervised model FARE-unsup and the semi-supervised model FARE-semisup. ACID \cite{15} is likewise an intrusion detection method based on adaptive clustering. In addition to the normalization operation for numerical features, it is necessary to be notified that in constructing the neighborhood matrix required for FARE, the 64 classifiers we build include 25 K-Means, 20 DBSCAN, 10 Birch, and 9 Minibatch K-Means.

\textbf{1) NSL-KDD:} We first conduct experiments on the NSL-KDD dataset \cite{9} and the experimental results are shown in Table \ref{tab3}. It is important to note that the value in the ``Category'' column (e.g., ``Normal'') is the precision of the category by the corresponding method. The results for each of these metrics are averaged over five folds. In addition, the ``Support'' row at the bottom of Table \ref{tab3} shows the number of samples in training set for each category. We observe that all these baseline methods perform worse than our Semi-WTC model at a low label size (1\% of the training set).

Specifically, Semi-WTC achieves 94.33\% accuracy and 92.73\% F1-score with only 1\% of the training set labels. It can also be seen that SOTA performance is achieved in several classes, such as ipsweep and normal. Among them, it is worth mentioning that Semi-WTC performs much better than the baseline methods FARE in detecting two categories such as ``warezclient'' and ``other''. The better performance is because Semi-WTC introduces a category weight factor through the weight consistency function, which can effectively balance the impact of long-tailed classes on the prediction of sparse classes. This experimental result also indicates that the clustering method has strong category sensitivity, and the classification accuracy is low when the sample distribution pattern is inconsistent (e.g., nmap) or the sample size of the category is small (e.g., other). In contrast, good results can be achieved for the opposite category. For example, ACID achieves 98.2\%  classification accuracy in class ``teardrop'', and FARE achieves 100\% classification accuracy in class ``smurf''. In addition, the rest of the traditional machine learning methods and VGGNet encounter greater limitations when the amount of labels is minimal (1\% of the training set) due to the difficulty of feature extraction, especially for SGD which achieves only 82.74\% classification accuracy. Our proposed Semi-WTC model achieves SOTA classification accuracy and is more suitable for network traffic datasets with label imbalance.

\textbf{2) CIC-IDS 2018:} To explore the generalization capability of Semi-WTC on different intrusion traffic datasets, we also conduct experiments on the CIC-IDS 2018 dataset \cite{10}, and the experimental results are shown in Table \ref{tab4}. Among them, the ``\emph{Time}'' column includes the time for reading data, pre-processing and training. The time is averaged over three experiments on the server. The rest of the experimental settings are the same as the NSL-KDD dataset.

Our Semi-WTC model achieves 97.66\% accuracy and 97.39\% F1-score with the shortest runtime of 64 s. The experiments show that
the Semi-WTC model achieves SOTA performance with more substantial generalization capability on different intrusion
traffic datasets compared to several other classification methods. In terms of running time, the FARE \cite{12} model takes the longest time, while the ACID \cite{15} model takes a shorter time but still has a longer running time than Semi-WTC. The running time of ACID is also lower than our proposed Semi-WTC model due to not using parallelization operation.
\vspace{-2 mm}
\begin{table}[!ht]
\vspace{-2mm}
    \setlength{\abovecaptionskip}{0pt}
    \setlength{\belowcaptionskip}{0pt}
	\LARGE
	\caption{Experimental results on the CIC-IDS 2018 dataset with 1\% labeled training data}
	\label{tab4}
	\tabcolsep 9pt 
	\renewcommand{\arraystretch}{1.2}
	\resizebox{\linewidth}{!}
	{
	\begin{tabular*}{\textwidth}{ccccc}
	\toprule
	  Model &Acc (\%)&Macro-F1 (\%) &TPR(\%)&Time (s)\\\hline
	  KMeans \cite{29} &67.09	&61.75	&65.01&3891 \\
	  RF \cite{30} &89.83	&89.17	&86.89&1922 \\
	  SVM \cite{31} &87.74 &85.89	&84.77&312 \\
	  LR / SGD \cite{32} &85.84	&83.66	&83.72&395 \\
	  LSTM \cite{33} &88.97	&87.13	&86.06&434 \\
	  VGGNet \cite{34} &91.84	&89.97	&89.88&531 \\
	  ACID \cite{15}  &92.58	&91.87	&90.53&347 \\
	  FARE-unsup \cite{12} &72.56	&68.32	&70.71&4378 \\
	  FARE-semisup \cite{12} &90.17	&89.51	&88.29&4512 \\
	  MixMatch \cite{44} &93.55	&92.29	&91.41&622 \\
	  Semi-WTC   &\textbf{97.66}	&\textbf{97.39}	&\textbf{95.60}&\textbf{64} \\
	\bottomrule
	\end{tabular*}
	}
\end{table}

\vspace{-3mm}
\subsubsection{\textbf{Performance Improvement with AAR}}
We investigate the performance improvement on our Semi-WTC model using AAR for the case of blurred distribution boundaries, in which 1\% of the data samples from the unseen class of the dataset are labeled by AAR every five epochs. The samples are then placed into the labeled training set to train the model. The experiments are conducted on the NSL-KDD dataset, and the results are shown in Table \ref{tab15}.

The experimental results indicate that the model performance fluctuates slightly when only 1\% of the unseen samples are expanded into the labeled dataset. However, in general, the model's performance improves after a certain level of training and achieves the performance without unseen samples. The fluctuations are that a small number of unseen core samples slightly affect the classification decision boundary of the model, resulting in the misclassification of samples near the decision boundary. However, after several learning rounds, the gains from learning the distribution of core sample data will offset the impact on the decision boundary. The distribution of data samples after multiple rounds of Active Adaption Resampling is shown in Fig. \ref{f6}.
\begin{table}[!ht]
\vspace{-4mm}
\setlength{\abovedisplayskip}{0pt}
\setlength{\belowdisplayskip}{0pt}
    \setlength{\abovecaptionskip}{0pt}
    \setlength{\belowcaptionskip}{0pt}
	\LARGE
	\caption{Experimental results when expanding 1\% unseen samples into labeled train set by AAR at different epochs}
	\label{tab15}
	\tabcolsep 20pt 
	\renewcommand{\arraystretch}{1.2}
	\resizebox{\linewidth}{!}
	{
	\begin{tabular*}{\textwidth}{cccc}
	\toprule
		Iteration &Acc (\%) &Macro-F1 (\%)&TPR (\%)\\\hline
	  	Epoch=0 (Initial) &87.81	&82.37&85.73\\
	  	Epoch=5  &87.49	&82.25&85.54\\
	  	Epoch=10  &88.96	&84.23&86.90\\
	  	Epoch=15 &90.79	&87.83&88.82\\
	  	Epoch=20  &92.78	&89.51&90.81\\
	  	Epoch=25  &93.93	&91.46&91.88\\
	  	Epoch=30 &{\textbf{94.62}}	&{\textbf{92.38}}&\textbf{92.64}\\
	  	Without Unseen  &94.33	&92.73 &92.37\\
	\bottomrule
	\end{tabular*}
	}
	\vspace{-1mm}
\end{table}
\begin{figure}[!ht]
    \vspace{-2mm}
    \setlength{\belowcaptionskip}{0pt}
    \centerline{\includegraphics[width=\columnwidth]{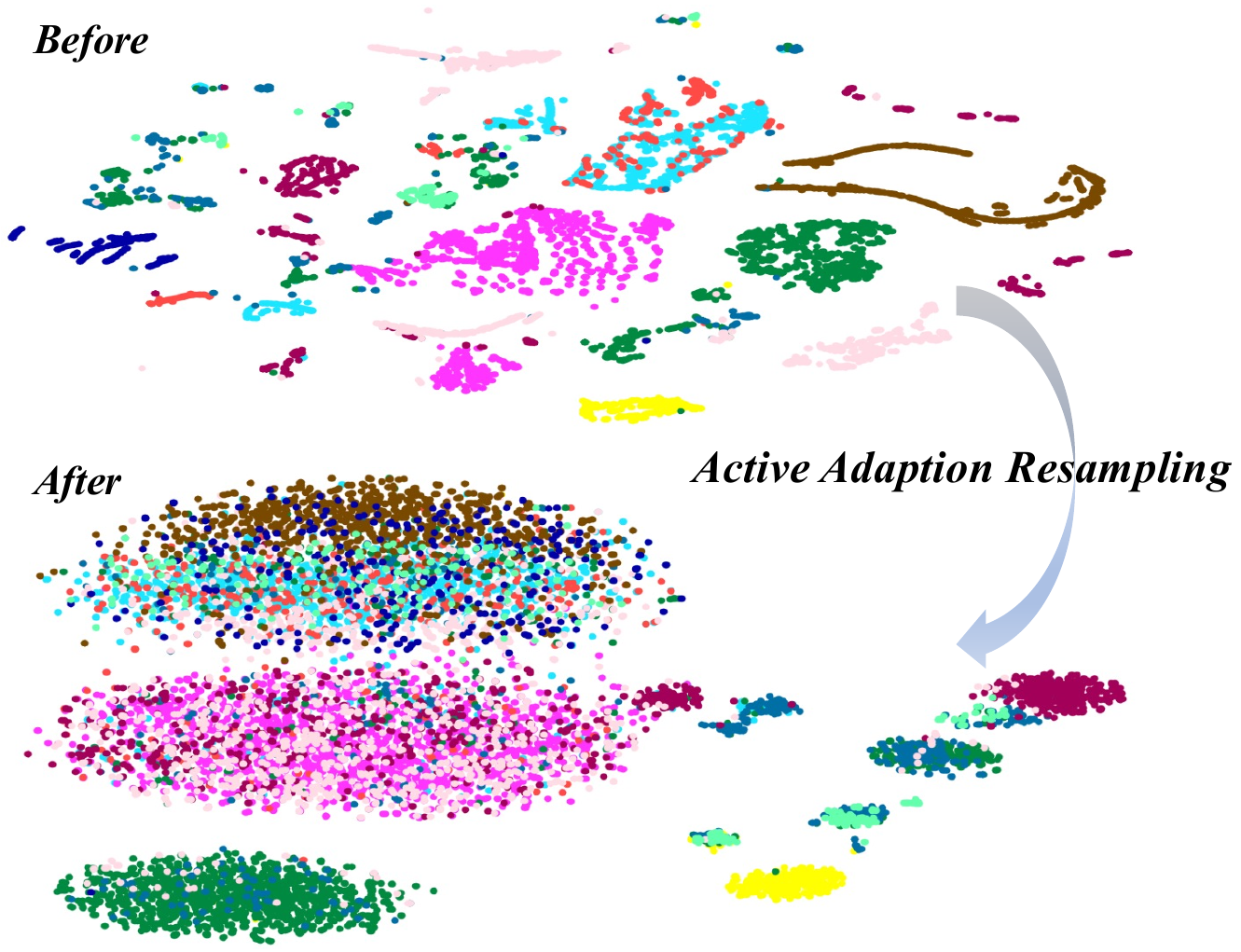}}
	\caption{Distribution of data after multiple active adaption resamplings. The core samples that characterize the class distribution are well resampled.}
	\label{f6}
	\vspace{-2mm}
\end{figure}

\subsubsection{\textbf{Comparison under Different Label Ratios}}
To further evaluate the improvement of the model in saving manual labeling, we conduct experiments on the CIC-IDS 2018 dataset with different ratios of labeled and unlabeled samples. We chose 30 times the data size of the previously used dataset to better reflect the practical effect, with a total data volume of 1,046,298 as shown in Table \ref{tab1}. We compare Semi-WTC with the SOTA FARE-semi, the semi-supervised method for fine-grained attack detection. 

The results are shown in Table \ref{tab2}, of which 1\%, 5\%, and 10\% of the initial training set are labeled, respectively. We can find that the proposed Semi-WTC outperforms FARE-semi\cite{12} in all cases. It can be seen that the classification accuracy and F1-score gradually improve with the increased labeled sample size. However, it is worth noting that our model achieves only 85.14\% classification accuracy and 79.53\% F1-score when labeling 0.5\% of the training set samples, but when the number of labeled data increases to 1\%, the classification accuracy and F1-score are significantly improved by 11.33\% and 14.16\%. However, the upward trend of classification accuracy and F1-score slows down as the number of labeled data increases. This change is because our proposed Semi-WTC model can capture the sample distribution of category data when only 1\% of labeled data is used. In contrast, 0.5\% of labeled samples is still not enough to support the model to learn the potential features of each category. Hence, the classification accuracy and other metrics improve significantly when the number of labeled data increases from 0.5\% to 1\%.
\vspace{-2 mm}
\begin{table}[!ht]
\vspace{-2mm}
    \setlength{\abovecaptionskip}{0pt}
    \setlength{\belowcaptionskip}{0pt}
	\LARGE
	\caption{Comparison of semi-supervised methods with different label ratios}
	\label{tab2}
	\tabcolsep 20pt 
	\renewcommand{\arraystretch}{1.2}
	\resizebox{\linewidth}{!}
	{
	\centering
	\begin{tabular*}{\textwidth}{cccc}
	\toprule
		Label Ratio&Method &Acc (\%)&Macro-F1 (\%)\\\hline
		\multirow{2}*{0.5\%} & Semi-WTC & \textbf{85.14} & \textbf{79.53}\\
		~ & FARE-semi\cite{12} & 80.25 & 71.37\\\hline
		\multirow{2}*{1\%} & Semi-WTC & \textbf{96.47} & \textbf{93.69}\\
		~ & FARE-semi\cite{12} & 89.76 & 88.20\\\hline
		\multirow{2}*{5\%} & Semi-WTC & \textbf{96.89} & \textbf{95.41}\\
		~ & FARE-semi\cite{12} & 91.86 & 90.04\\\hline
		\multirow{2}*{10\%} & Semi-WTC & \textbf{97.04} & \textbf{96.11}\\
		~ & FARE-semi\cite{12} & 92.31 & 91.22\\
	\bottomrule
\end{tabular*}
}
\vspace{-2 mm}
\end{table}

\subsubsection{\textbf{Influence of Hyper-parameters}}
\begin{figure}[!ht]
    \setlength{\belowcaptionskip}{0pt}
    \vspace{-2mm}
	\centerline{\includegraphics[width=\columnwidth]{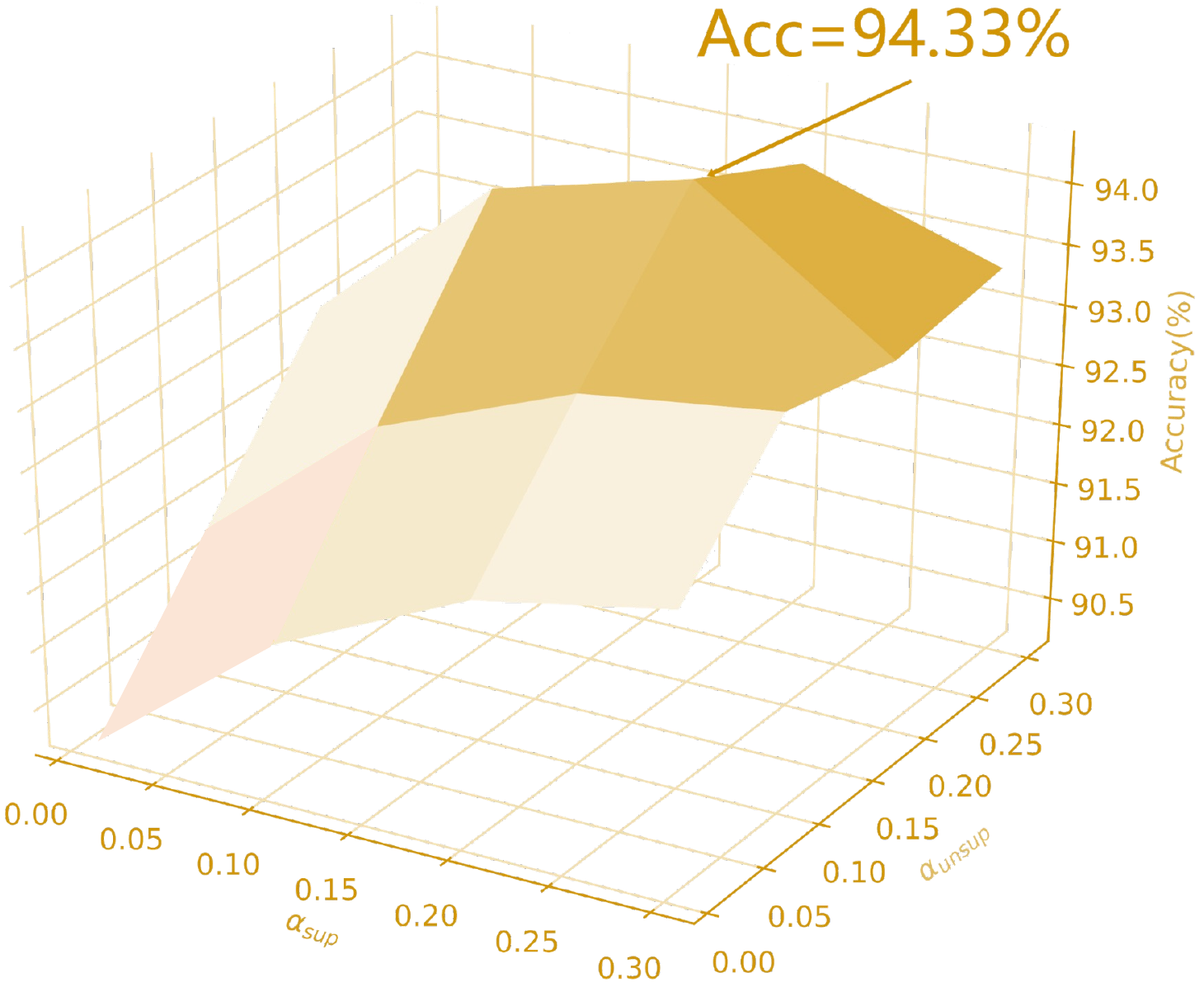}}
	\caption{Experimental results of different $\alpha_{sup}$ and $\alpha_{unsup}$}
	\label{f5}
	\vspace{-3mm}
\end{figure}

We conduct experiments for the model's hyperparameters, where we choose the MSE loss weighting factor. And the weighting factors $\alpha_{sup}$ and $\alpha_{unsup}$ for four experiments are 0.01 , 0.1, 0.2 and 0.3, respectively. We evaluate the proposed model's influence on the hyperparameters, i.e., the weighting factors for the MSE loss. The results of setting weighting factors $\alpha_{sup}$ and $\alpha_{unsup}$ to 0.01 , 0.1, 0.2 and 0.3, respectively, are shown in Fig.~\ref{f5}.

According to the experimental results, the Semi-WTC model achieves optimal performance with 94.33\% classification accuracy at $\alpha_{unsup}=0.2$ and $\alpha_{sup}=0.2$, while it achieves sub-optimal performance with 93.92\% accuracy at $\alpha_{unsup}=0.1$ and $\alpha_{sup}=0.2$. From the trend, when $\alpha_{unsup}$ and $\alpha_{sup}$ are relatively balanced, a better classification is often achieved. We believe this is because the Semi-WTC model needs to extract similar features for both labeled and unlabeled data, and thus requires a relatively balanced weighting factor to balance the two losses. Secondly, we also find an interesting point that the model performance tends to be better when $\alpha_{sup}>\alpha_{unsup}$ than when $\alpha_{sup}<\alpha_{unsup}$. This trend may be because the model needs to learn more about the data distribution of the labeled samples. Thus when the loss weight of the unlabeled part is higher than the labeled part, the model is affected by the pseudo-label distribution from the data in the unlabeled part. The pseudo-label distribution is not accurate enough compared to the real label distribution, which will lead to a decrease in the classification accuracy.
\vspace{-3mm}
\subsubsection{\textbf{Ablation Experiments}}
Ablation experiments aim to investigate the robustness of our model against fewer labels. We conduct ablation experiments on the NSL-KDD dataset to evaluate the effects of network structure, sample size, and loss weight factor. To this end, we design several types of ablation experiments to study MLP (RB-MLP (w/o)) and RB-MLP with and without \textbf{W}eight-\textbf{T}ask \textbf{C}onsistency (WTC). The experimental results are shown in Table \ref{tab5}. 

\begin{table}[!ht]
\vspace{-2mm}
    \setlength{\abovecaptionskip}{0pt}
    \setlength{\belowcaptionskip}{0pt}
	\centering
	\LARGE
	\caption{Ablation experiments about RB-MLP and WTC on the NSL-KDD dataset with 1\% labeled training data}
	\label{tab5}
	\tabcolsep 25pt 
	\renewcommand{\arraystretch}{1.2}
	\resizebox{\linewidth}{!}
	{
	\begin{tabular*}{\textwidth}{ccccc}
	\toprule
	  RB-MLP &(w/o)	&\checkmark &(w/o) &\checkmark\\
	  WTC &(w/o) &(w/o) &\checkmark &\checkmark\\\hline
	  Acc (\%)&91.34 &92.30 &91.93	&94.33\\
	  Macro-F1 (\%)&88.40 &89.85 &89.07	&92.73\\
	\bottomrule
	\end{tabular*}
	}
\end{table}

It can be seen that the introduction of RB-MLP and WTC both lead to different degrees of performance improvement. Introducing residual connectivity in the traditional MLP results in a higher performance improvement than WTC. Changing the MLP to RB-MLP improves the classification accuracy by 0.96\% and 2.4\% without and with Weight-Task Consistency, respectively, while introducing WTC
improves the accuracy by 0.59\% and 2.03\% without and with RB-MLP, respectively. We believe this is because RB-MLP fuses the high-level and underlying information of the sample, which can extract more comprehensive features and thus achieve high-precision classification.

We also conduct ablation experiments for Semi-WTC to investigate the specific performance of CUM. The experimental results are shown in Table \ref{tab7-2}. From the experimental results, it can be seen that the improvements of CUM are practical. The introduction of CUM leads to performance improvement on all the datasets. The most significant improvement in the accuracy metric was obtained on the AndMal 2020 dataset, with a 1.03\% improvement. Among them, by introducing the CUM, the Marco-F1 score and TPR score improve by up to 1.14\% and 0.87\%, respectively, compared to Semi-WTC without CUM. Furthermore, by introducing CUM, model performance improvement is ensured by strengthening important features and suppressing irrelevant features during the training phase. It proves that CUM will minimize the uncertainty of the model due to feature diversity and enhance the model's effectiveness.

\vspace{-2mm}
\begin{table}[!ht]
    \vspace{-2mm}
    \setlength{\abovecaptionskip}{0pt}
    \setlength{\belowcaptionskip}{0pt}
	\LARGE
	\caption{Ablation experiments about CUM (*) on the NSL-KDD, CIC-IDS 2018 and AndMal 2020 datasets with 1\% labeled training data}
	\label{tab7-2}
	\tabcolsep 10pt 
	\renewcommand{\arraystretch}{1.2}
	\resizebox{\linewidth}{!}
	{
	\centering
	\begin{tabular*}{\textwidth}{ccccc}
	\toprule
		Datasets&Method &Acc (\%)&Macro-F1 (\%)&TPR (\%)\\\hline
		\multirow{2}*{NSL-KDD} & Semi-WTC* & \textbf{95.16} & \textbf{92.98}&\textbf{92.55}\\
		~ & Semi-WTC & 94.33 & 92.73 &92.37\\\hline
		\multirow{2}*{CIC-IDS 2018} & Semi-WTC* & \textbf{98.02} & \textbf{97.87}&\textbf{96.18}\\
		~ & Semi-WTC & 97.66& 97.39& 95.60\\\hline
		\multirow{2}*{AndMal 2020} & Semi-WTC* & \textbf{95.09} & \textbf{94.86}&\textbf{92.84}\\
		~ & Semi-WTC & 94.06& 93.72& 91.97\\
	\bottomrule
\end{tabular*}
}
\vspace{-2 mm}
\end{table}

\vspace{-2mm}
\subsection{\textbf{Experiments Simulating Realistic Situations}}
We conduct various experiments for realistic situations, including unseen classes, which means the test set includes categories that are not in the training set. Other realistic situations include mislabelling, imbalance, and big datasets. In this section, FARE refers to the semi-supervised version of FARE unless otherwise specified.
\vspace{-4mm}
\subsubsection{\textbf{Handling unseen classes}}
The unseen class experiment focuses on verifying the end-to-end performance of the Semi-WTC model under settings with unseen classes. More specifically, we simulate the scenario where several classes that do not exist in the training set appear in the test set. Here we conduct experiments on the NSL-KDD \cite{9} dataset, where we first count all the classes in the training set and the test set, respectively, for a total of $N$. After that, $N/10$ of the classes are randomly selected for the training set. All samples of these classes are moved into the test set to achieve the setting for unseen classes. Then the pre-processing described in Section \ref{prepro} is performed. The experimental results are shown in Table \ref{tab7}.

\vspace{-0.5mm}
\begin{table}[!ht]
\vspace{-2mm}
    \setlength{\abovecaptionskip}{0pt}
    \setlength{\belowcaptionskip}{0pt}
	\LARGE
	\caption{Unseen-class experimental results on NSL-KDD dataset with 1\% labeled training data}
	\label{tab7}
	\tabcolsep 38pt 
	\renewcommand{\arraystretch}{1.2}
	\resizebox{\linewidth}{!}
	{
	\begin{tabular*}{\textwidth}{ccc}
	\toprule
	  Model &Acc (\%)&Macro-F1 (\%)\\\hline
	  FARE (unseen) \cite{12}	&86.19	&80.23\\
	  FARE\cite{12}&88.21 	&82.57\\
	  ACID (unseen) \cite{15}	&87.46	&85.18\\
	  ACID \cite{15}	&91.05 	&88.91\\
	  Semi-WTC (unseen)	&{\textbf{92.84}}	&{\textbf{91.54}}\\
	  Semi-WTC	&\textbf{94.33}	&\textbf{92.73}\\
	\bottomrule
	\end{tabular*}
	}
\end{table}
\begin{table*}[!ht]
    \setlength{\abovecaptionskip}{0pt}
    \setlength{\belowcaptionskip}{0pt}
	\small
	\caption{Imbalance experimental results on the CIC-IDS 2018 big dataset with 1\% labeled training data and metric is accuracy}
	\label{tab9}
	\tabcolsep 2.5pt 
	\renewcommand{\arraystretch}{1.2}
	\resizebox{\linewidth}{!}
	{
	\begin{tabular*}{\textwidth}{cccccccccc}
	\toprule
		Category &KMeans\cite{29}	&RF\cite{30}	&SVM\cite{31}	&LR/SGD\cite{32}	&LSTM\cite{33}	&VGGNet\cite{34} 	&ACID\cite{15}	&FARE un/semi\cite{12}	&Semi-WTC \\\hline
	  other & 38.9	&46.4	&50.6	&44.1	&45.7	&66.7	&72.3	&40.5 / 45.7	&\textbf{77.1}\\
	  other$_{im}$ &34.1	&43.3	&46.7	&40.2	&41.6	&63.1	&65.9	&38.4 / 42.6	&{\textbf{74.2}}\\
	\bottomrule
	\end{tabular*}
	}
	\vspace{-1mm}
\end{table*}

It can be seen that the overall classification accuracy and F1-scores of all models decrease in the presence of unseen samples, which is because the models cannot effectively learn the data distribution of the unseen samples. However, it can also be seen that in the case of unseen samples, the classification accuracy of the Semi-WTC model is 92.84\%, and the F1-score is 91.54\%, where the decreases are significantly lower than those of the ACID and FARE models, and can be higher than the normal classification (without unseen classes) accuracy and F1-score of ACID and FARE models. These results show that the Semi-WTC model can effectively generalize the classification to unseen samples. Because the introduction of category weight coefficients $\delta_{i}$ can effectively control the influence of head classes and then increase the recognition ability for tail classes to have better generalization ability for unseen samples.

\vspace{-2mm}
\subsubsection{\textbf{Handling imbalanced classes}}
In the imbalance experiment, 10\% of the original data (67 samples) in the class ``other'' are utilized for training to verify our model's generalization to the imbalanced classes. The experimental results are shown in Table \ref{tab9}. The accuracy of the Semi-WTC model for the class ``other'' with a much smaller sample size only decreases by 2.9\%, which is much lower than the decrease of 6.4\% of the second-best model ACID. This decrease indicates that the Semi-WTC model also has a particular generalization effect on highly imbalanced data.
\vspace{-3mm}
\subsubsection{\textbf{Handling mislabeled samples}}
This experiment aims to evaluate the performance of Semi-WTC with partially mislabeled training samples. In this case, we randomly swap 10\% of the samples in the labeled training set with another 10\% of the sample labels to achieve the case of artificial sample mislabeling. The results are shown in Table \ref{tab8}. It is an end-to-end evaluation of Semi-WTC in a partially mislabeled sample setting.

The experimental results show that Semi-WTC has a higher decrease in accuracy and F1-score than ACID and FARE in the case of wrong labels, reaching 2.3\% and 1.5\%, respectively, because of the different internal mechanisms of the models. Semi-WTC essentially belongs to the perceptron model, while ACID and FARE belong to the clustering model, so the importance of each sample is similar in Semi-WTC. In contrast, for clustering, the samples at the center of the clusters tend to be more critical, which leads to the slightly worse robustness of Semi-WTC to mislabelling than ACID and FARE. However, it is worth stating that due to the higher base classification accuracy, the classification accuracy of Semi-WTC is still higher than the other two models even if mislabelled training samples exist.

\begin{table}[!ht]
    \vspace{-2mm}
    \setlength{\abovecaptionskip}{0pt}
    \setlength{\belowcaptionskip}{0pt}
	\LARGE
	\caption{Mislabeled experimental results on the NSL-KDD dataset with 1\% labeled training data}
	\label{tab8}
	\tabcolsep 32pt 
	\renewcommand{\arraystretch}{1.2}
	\resizebox{\linewidth}{!}
	{
	\begin{tabular*}{\textwidth}{ccc}
	\toprule
	  Model &Acc (\%)&Macro-F1 (\%)\\\hline
	  FARE (mislabeled) \cite{12}	&88.07	&82.48\\
	  FARE \cite{12}	&88.21	&82.57\\
	  ACID (mislabeled) \cite{15}	&90.62	&88.65\\
	  ACID \cite{15}	&91.05	&88.91\\
	  Semi-WTC (mislabeled)	&{\textbf{92.03}}	&{\textbf{91.29}}\\
	  Semi-WTC	&\textbf{94.33}	&\textbf{92.73}\\
	\bottomrule
	\end{tabular*}
	}
	\vspace{-2mm}
\end{table}

\subsubsection{\textbf{Handling large-scale dataset}}
To explore the performance of the Semi-WTC model with large-scale datasets, we conduct additional experiments on the CIC-IDS 2018 big dataset \cite{10}, which has a total data volume of 1,046,298 items. The data size is 30 times that of the previously used dataset for our experiments, shown in Table \ref{tab11}. The division of the dataset is shown in Table \ref{tab1}. The comparison experiments are conducted on the two models (Semi-WTC, ACID) that performed best and second-best on the small CIC-IDS 2018 dataset. The specific experimental results are shown in Table \ref{tab11}.

\begin{table}[!ht]
\vspace{-2mm}
    \setlength{\abovecaptionskip}{0pt}
    \setlength{\belowcaptionskip}{0pt}
	\LARGE
	\caption{Experimental results on the CIC-IDS 2018 big dataset with 1\% labeled training data}
	\label{tab11}
	\tabcolsep 20pt 
	\renewcommand{\arraystretch}{1.2}
	\resizebox{\linewidth}{!}
	{
	\begin{tabular*}{\textwidth}{cccc}
	\toprule
	  Model &Acc (\%)	&Macro-F1 (\%)	&Time (s)\\\hline
	  ACID (big)\cite{15}	&90.31	&86.45	&5275\\
	  ACID (small)\cite{15}	&92.58	&91.87	&357\\
	  FARE (big)\cite{12}	&89.76	&88.20	&59144\\
	  FARE (small)\cite{12}	&90.17	&89.51	&4522\\
	  Semi-WTC (big)	&{\textbf{97.02}}	&{\textbf{94.91}}	&{\textbf{1319}}\\
	  Semi-WTC (small)	&\textbf{97.66}	&\textbf{97.39}	&\textbf{74}\\
	\bottomrule
	\end{tabular*}
    }
    \vspace{-2mm}
\end{table}

It can be seen that the performance of both models degrades on the big dataset. This is because the data distribution is more diverse in the big dataset, making learning for models with limited parameters difficult. However, Semi-WTC still performs much more efficiently because the multilayer perceptron-based model has fewer parameters than ACID, and Semi-WTC has fewer network layers that can effectively identify different features.

\subsubsection{\textbf{Practical Deployment Experiments}}
After the controlled experiments, we also validate the Semi-WTC model in real deployments, as shown in Fig. \ref{f7}. We first briefly describe the collection and cleaning of the actual deployment data in the following. The dataset contains traffic data collected from IoT devices for about four hours daily. Our dataset includes normal traffic generated by IoT devices and injected attack traffic together with responses. Our judgment criteria for attack traffic is that communication between non-gateways and intranet addresses in IoT device addresses or communication with IoT devices are injected anomalous traffic.

Specifically, we extract $x$ normal traffic samples and $y$ abnormal traffic samples (including attack categories such as port scan, DoS, and DDoS), respectively, where each traffic sample is represented as a \emph{20}-dimensional feature vector. The feature vectors are encoded using their internal feature engineering methods, including information such as the port number, message length, and flag bits. 
\begin{table}[!ht]
\vspace{-2mm}
    \setlength{\abovecaptionskip}{0pt}
    \setlength{\belowcaptionskip}{0pt}
	\LARGE
	\caption{Experimental results of the Semi-WTC model in real deployment experiment with 1\% labeled training data}
	\label{tab16}
	\setlength{\tabcolsep}{27pt}
	\renewcommand{\arraystretch}{1.2}
	\resizebox{\linewidth}{!}
	{
	\begin{tabular*}{\textwidth}{cccc}
	\toprule
	  Model &Acc (\%) &F1 (\%) &FPR (\%) \\\hline
	  FARE\cite{12} (reality) &90.59 &90.06 &4.66\\
	  ACID\cite{15} (reality) &87.46 &87.20 &11.02\\
	  Semi-WTC (reality) &\textbf{95.05} &\textbf{94.95} &\textbf{3.44}\\
	\bottomrule
	\end{tabular*}
	}
\end{table}
The methods for dividing the dataset are referred to Table \ref{tab1}, which is divided into the labeled training set, unlabeled training set, validation set, and test set. We apply FARE, ACID, and Semi-WTC on this dataset, respectively. And the results are shown in Table \ref{tab16}. We observe that the Semi-WTC model still has good generalization performance for real network traffic. It achieves a classification accuracy of 95.05\%, which is 7.59\% and 4.46\% higher than the ACID and FARE models, respectively. The accuracy also indicates that our model has better robustness to the data distribution of the imbalanced class. The dataset is directly connected to the Internet, and there may be anomalies, which will impact. Furthermore, We believe that it is the main reason the overall effect of the actual environment deployment is not good. 
\begin{figure}[!ht]
    \setlength{\belowcaptionskip}{-2pt}
    \centering
	\centerline{\includegraphics[width=0.45\textwidth]{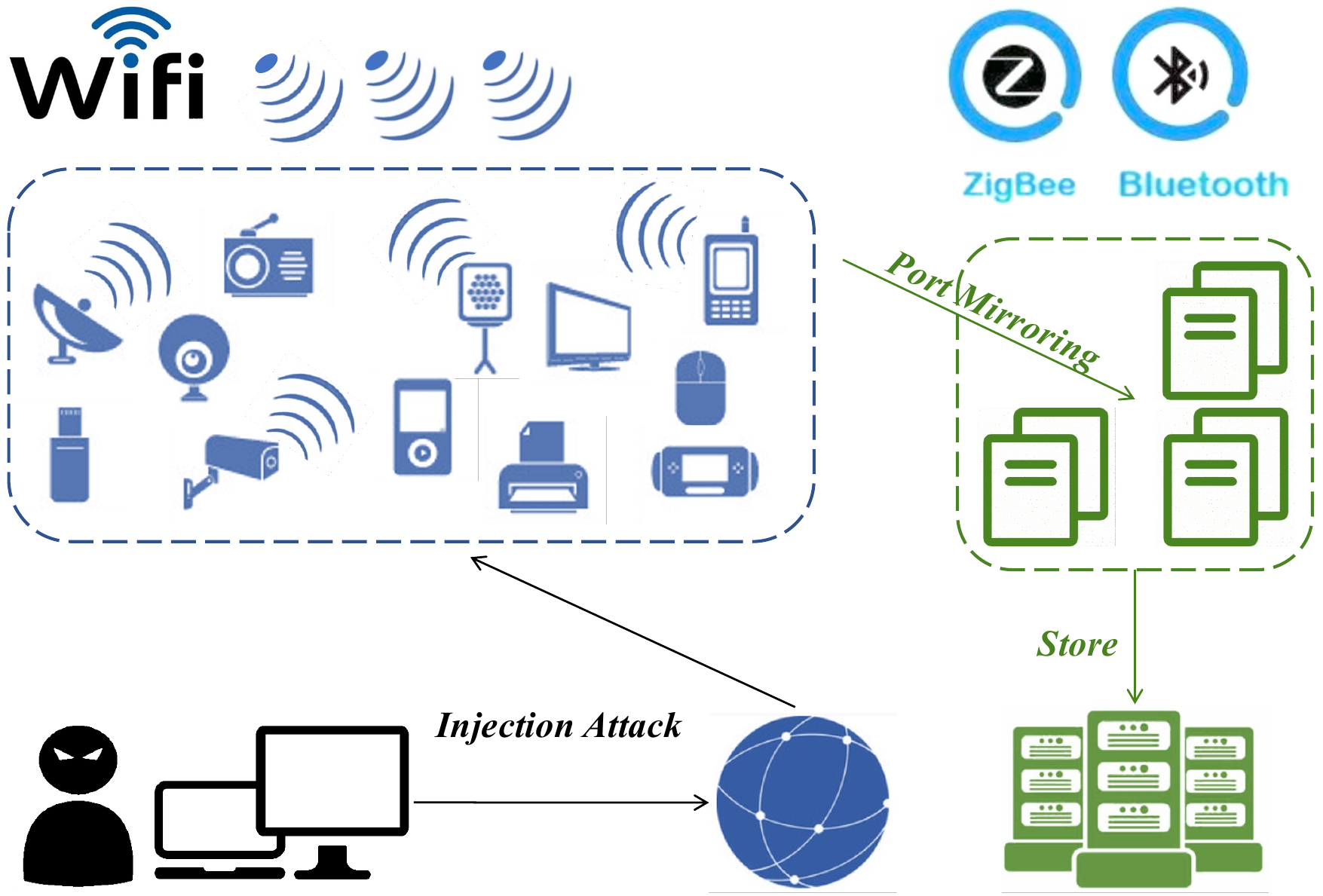}}
	\caption{Deployment diagram in a real campus network environment}
	\label{f7}
	\setlength{\abovecaptionskip}{-2pt}
	\vspace{-1 mm}
\end{figure}
\vspace{-4mm}

\subsection{\textbf{Robustness Experiments for Semi-WTC}}
\subsubsection{\textbf{Poisoning Attack Experiments}}

In semi-supervised learning, Carlini et al. proposed \cite{43} that the model performance will be significantly affected by poisoning attacks on unlabeled datasets. This is because the usual semi-supervised mechanism obtains pseudo labels by predicting unlabeled data. At the same time, we can inject misleading features into the unlabeled dataset, leading the model to mislabel the unlabeled data with high probability. So, during the training phase, poisoning attack experiments are conducted on our proposed Semi-WTC, FARE, and ACID following the experimental settings in \cite{43}. 


We employ the poisoning attack method in \cite{42}, which uses the TTL value in the traffic IP header as a channel for backdoor signals. It is assumed that the TTL field remains constant for all packets in benign traffic, and this assumption holds for the dataset we use, where less than 1\textperthousand of the normal traffic data in the CIC-IDS 2018 dataset exhibits a non-zero standard deviation in the TTL value. Therefore, the CIC-IDS 2018 dataset performs poisoning attacks on the model by varying the TTL of the traffic samples in the unlabeled dataset. For example, the poisoning is achieved by increasing the TTL of the first packet by one if its TTL is less than 128 and decreasing it by one if it is greater than or equal to 128. 
\vspace{-1mm}
\begin{table}[!ht]
\vspace{-2mm}
    \setlength{\abovecaptionskip}{0pt}
    \setlength{\belowcaptionskip}{0pt}
	\LARGE
	\caption{Poisoning attack experimental results on the CIC-IDS 2018 datasetat with 1\% labeled training data}
	\label{tab12}
	\tabcolsep 38pt 
	\renewcommand{\arraystretch}{1.2}
	\resizebox{\linewidth}{!}
	{
	\begin{tabular*}{\textwidth}{ccc}
	\toprule
	  Model &Acc (\%)&Macro-F1 (\%)\\\hline
	  FARE\cite{12} (poison) &88.09 &82.14\\
	  ACID\cite{15} (poison) &89.94 &87.13\\
	  Semi-WTC (poison) &\textbf{93.10} &\textbf{91.22}\\
	\bottomrule
	\end{tabular*}
	}
	\vspace{-1mm}
\end{table}

The experimental results are shown in Table \ref{tab12}. We observe that all models can maintain high classification accuracy even in poisoning attacks. Meanwhile, our model can maintain a 3\% to 5\% accuracy lead compared with the rest of the methods. This is because introducing the RPM loop structure in the Semi-WTC model can repeatedly learn the information of data labels and the potential information inside the data, which has better robustness against poisoning attacks.
\vspace{-2mm}
\subsubsection{\textbf{Adversarial Attack Experiments}}
During the testing phase, following the experimental settings and the adversarial attack sample generation method of IDS-MLP \cite{35}, adversarial attack experiments are conducted on our proposed Semi-WTC, FARE, ACID, and the IDS-MLP \cite{35}. IDS-MLP is a machine learning approach for intrusion detection using a Multilayer Perceptron (MLP) network, which can significantly reduce the accuracy of the IDS. Our adversarial attack relies on knowing the parameters used in the training model instead of the training data. Overall, our adversarial attack design is mainly based on the Jacobian-based Saliency Map Attack (JSMA) \cite{36} to create adversarial test samples. JSMA is based on the saliency map approach to generate adversarial samples $\tilde{X}=X+\sigma$, which typically has only a tiny perturbation $\sigma$, and thus may deceive the models. The experiments are conducted on the CIC-IDS 2018 dataset, and the results are shown in Table \ref{tab13}.
\vspace{-5mm}
\begin{table}[!ht]
\vspace{-2mm}
    \setlength{\abovecaptionskip}{0pt}
    \setlength{\belowcaptionskip}{0pt}
	\LARGE
	\caption{Adversarial attack experimental results on the CIC-IDS 2018 datasetat with 1\% labeled training data}
	\label{tab13}
	\tabcolsep 25pt 
	\renewcommand{\arraystretch}{1.2}
	\resizebox{\linewidth}{!}
	{
	\begin{tabular*}{\textwidth}{ccc}
	\toprule
		Model &Acc (\%)&Macro-F1 (\%)\\\hline
		FARE (adversarial)\cite{12} &78.65	&75.43\\
		ACID (adversarial)\cite{15} &71.32	&68.18\\
		IDS-MLP (adversarial)\cite{35} &75.96	&73.21\\
		Semi-WTC (adversarial) &\textbf{84.83}	&\textbf{81.79}\\
		Semi-WTC (normal) &97.66	&97.39\\
	\bottomrule
	\end{tabular*}
	}
	\vspace{-2mm}
\end{table}

The experimental results indicate that all models suffer from different degrees of performance loss when encountering well-designed adversarial attacks. Take our Semi-WTC as an example. Its classification accuracy decreases by 12.83\%, and the F1-score decreases by 15.6\%, but it still achieves the best performance compared to other methods. This is due to introducing a dual branch structure in Semi-WTC, which can extract semantic information of features for labeled and unlabeled data. Compared with the IDS-MLP model, which is also based on MLP/perceptron, Semi-WTC obtains better performance.
\vspace{-3mm}
\subsection{\textbf{Applications of Semi-WTC in Other Security Areas}}
This section presents experiments with the Semi-WTC on other security domains. The AndMal 2020 Dataset \cite{27}\cite{28} is taken as an example. The experimental results are shown in Table \ref{tab14}. The samples in this dataset have a much larger feature dimension, 9,441 dimensions in total, and therefore the running time of each model is longer. Since the FARE \cite{12} does not use the original features directly but uses the neighborhood matrix as a second representation of the sample features, its running time on the AndMal 2020 dataset is less different from that on the CIC-IDS 2018 dataset.
\begin{table}[!ht]
\vspace{-2mm}
    \setlength{\abovecaptionskip}{0pt}
    \setlength{\belowcaptionskip}{0pt}
	\LARGE
	\caption{Experimental results on the AndMal 2020 datasetat with 1\% labeled training data}
	\label{tab14}
	\tabcolsep 8pt 
	\renewcommand{\arraystretch}{1.2}
	\resizebox{\linewidth}{!}
	{
	\begin{tabular*}{\textwidth}{ccccc}
	\toprule
		Model &Acc (\%) &Macro-F1 (\%)&TPR (\%) &Time (s)\\\hline
		KMeans \cite{29} &70.15	&65.26 &	67.53&4278\\
		RF \cite{30} &90.89	&90.43 &	88.22&3509\\
		SVM \cite{31} &88.93 &86.78 &	86.30&617\\
		LR/SGD \cite{32} &87.52	&86.22 &	85.11&759\\
		LSTM \cite{33} &87.88	&87.35 &	85.46&843\\
		VGGNet \cite{34} &89.21	&88.67 &	86.54&1046\\
		ACID \cite{15} &90.87	&90.55 &	88.39&627\\
		FARE-unsup \cite{12} &74.61	&70.12 &72.18	&4710\\
		FARE-semisup \cite{12} &91.13	&90.79 &88.61	&4847\\
		MixMatch \cite{44} &90.78	&90.39	&88.36&835 \\
		Semi-WTC   &\textbf{94.06}	&\textbf{93.72} &\textbf{91.97}	&\textbf{146}\\
	\bottomrule
	\end{tabular*}
	}
	\vspace{-1mm}
\end{table}
In addition, the ACID \cite{15} relies on multiple kernel networks to learn the optimal embedding of data samples and needs to identify the clustering centers accurately. Hence, its performance in high-dimensional datasets is not as good as that of FARE-semisup \cite{12}. Overall, on the AndMal 2020 dataset \cite{27}\cite{28}, our proposed Semi-WTC achieves SOTA performance in all three aspects, including accuracy, F1-score, and running time. The improvement of its classification accuracy even reaches 24\% compared to the K-Means \cite{29} algorithm.
\vspace{-4mm}
\section{Discussions}
\textbf{Generality of the Framework.}
Essentially, our proposed Semi-WTC approach and the RB-MLP structure are not interdependent; The Semi-WTC framework is a general semi-supervised framework that can be combined with any neural network structure. We choose the fully connected layer with residual connections as the basis for network construction because of its good fault tolerance and nonlinear global action, representing the distribution of unseen data. In the future, we will try more network structures to explore the performance of our framework for different tasks.

\textbf{Limitations and Future Works.}
There are some limitations of our work. First, we only study the model's generalization capability on malware analysis and the long-tail effect on labeled data. We can conduct experiments for more domains in the future to explore the generalization ability of the model. Second, we conduct experiments on unseen classes, imbalanced classes, mislabeled samples, large-scale datasets, and poisoning attacks to simulate realistic environments as much as possible due to space limitations. We also deploy servers to monitor normal traffic and simulate injection attacks. However, in realistic environments, unseen classes and misclassifications may often co-occur, so more comprehensive experiments need to be conducted in the future to analyze them. We will conduct further experiments on other practical attacks, such as problem-space attacks \cite{r3pierazzi2020intriguing} \cite{r4han2021evaluating} in the future. Finally, due to the limitation of the number of layers of the encoding network, it is difficult to learn on large data sets with more diverse data distribution. The reason is that increasing the number of hidden layers of RB-MLP may lead to overfitting for a small amount of labeled data. In future work, we intend to explore other semi-supervised mechanisms that can balance the effects and long-tail effects on unlabeled data.

\vspace{-3mm}
\section{Conclusion}
In this paper, we propose a new semi-supervised framework Semi-WTC, which uses a two-branch structure to incorporate information from both supervised and unsupervised parts. The Semi-WTC includes RB-MLP (ResBMLP), RPM (Recurrent Prototype Module), and WTC (Weight-Task Consistency) modules. The RB-MLP and RPM are combined to tackle low-quality labels, and WTC is employed to rebalance categories. To label training samples that are more representative of the actual distribution and improve the model's generalization performance, we propose the AAR (Active Adaption Resampling) method. The target objects of the method are core samples that reflect the characteristics of the category distribution instead of the classification decision boundaries usually chosen by traditional active learning methods.

In summary, our Semi-WTC can well solve the problem of data imbalance and is resistant to adversarial attacks and poisoning attacks. We believe the proposed method could significantly reduce the burden of acquiring sufficient labels and improve the generalizability of the learned model to adapt to various imbalanced data distributions. Furthermore, thus promote future research into related areas. Meanwhile, several generic plug-and-play modules included in Semi-WTC are easily reusable in other models. By conducting experiments on three datasets, our semi-supervised framework outperforms existing other SOTA algorithms and can improve robustness and generality. Currently, our approach is only practiced in intrusion detection, network traffic monitoring, and malware monitoring and lacks application in more situations. In the future, we hope our model will be applied in more attack detection-related fields and scenarios.
\vspace{-4 mm}

%

\ifCLASSOPTIONcaptionsoff
  \newpage
\fi



%
\vspace{-1mm}
\bibliographystyle{ieeetr}
\bibliography{bare_jrnl_compsoc}

%






\vspace{-14mm}
\begin{IEEEbiography}[{\includegraphics[width=1in]{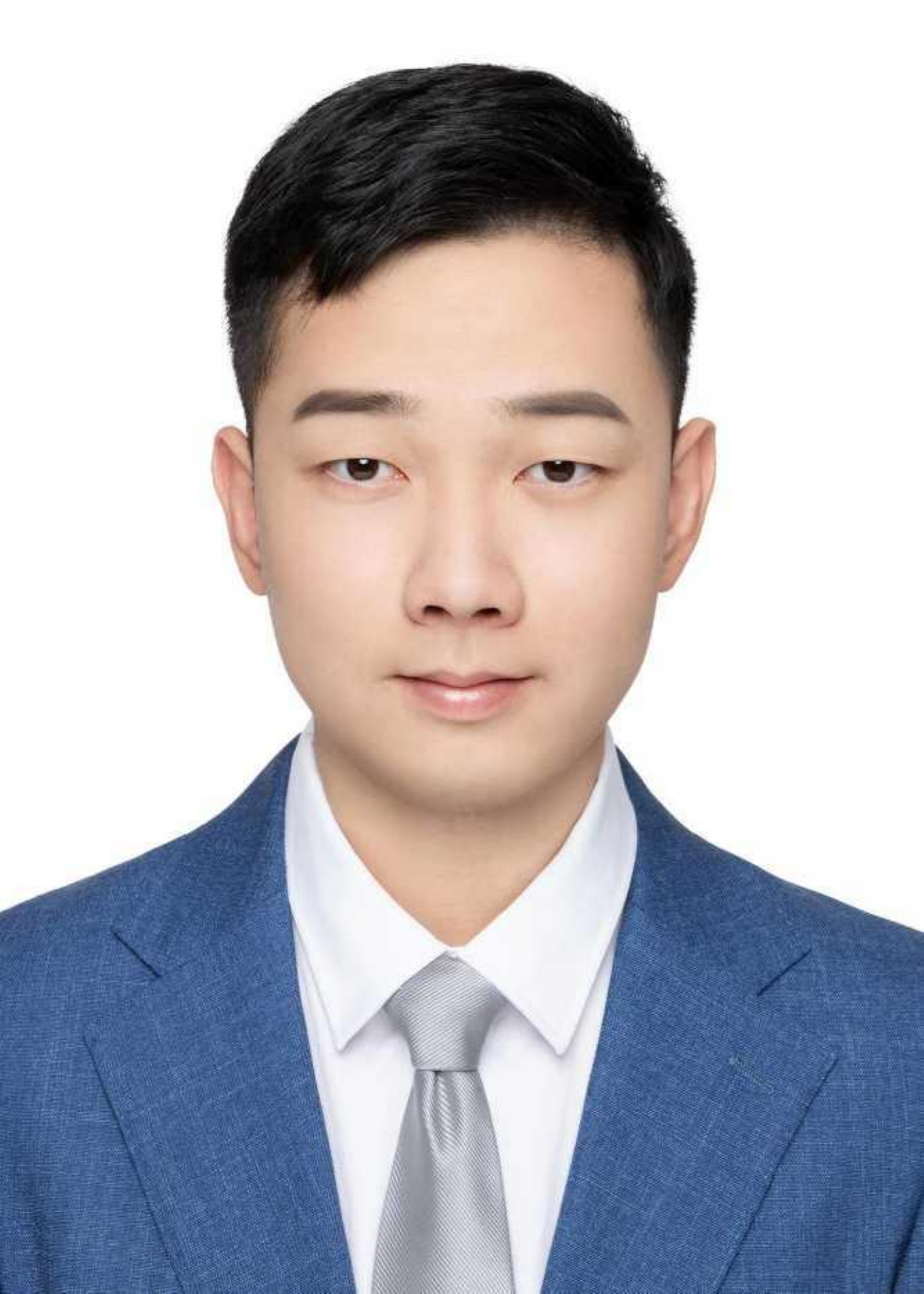}}]{Zihan Li} received the B.E. degree from Xiamen University. His research interests include network intrusion detection, pattern recognition and machine learning.
\end{IEEEbiography}

\vspace{-12mm}

\begin{IEEEbiography}[{\includegraphics[width=1in]{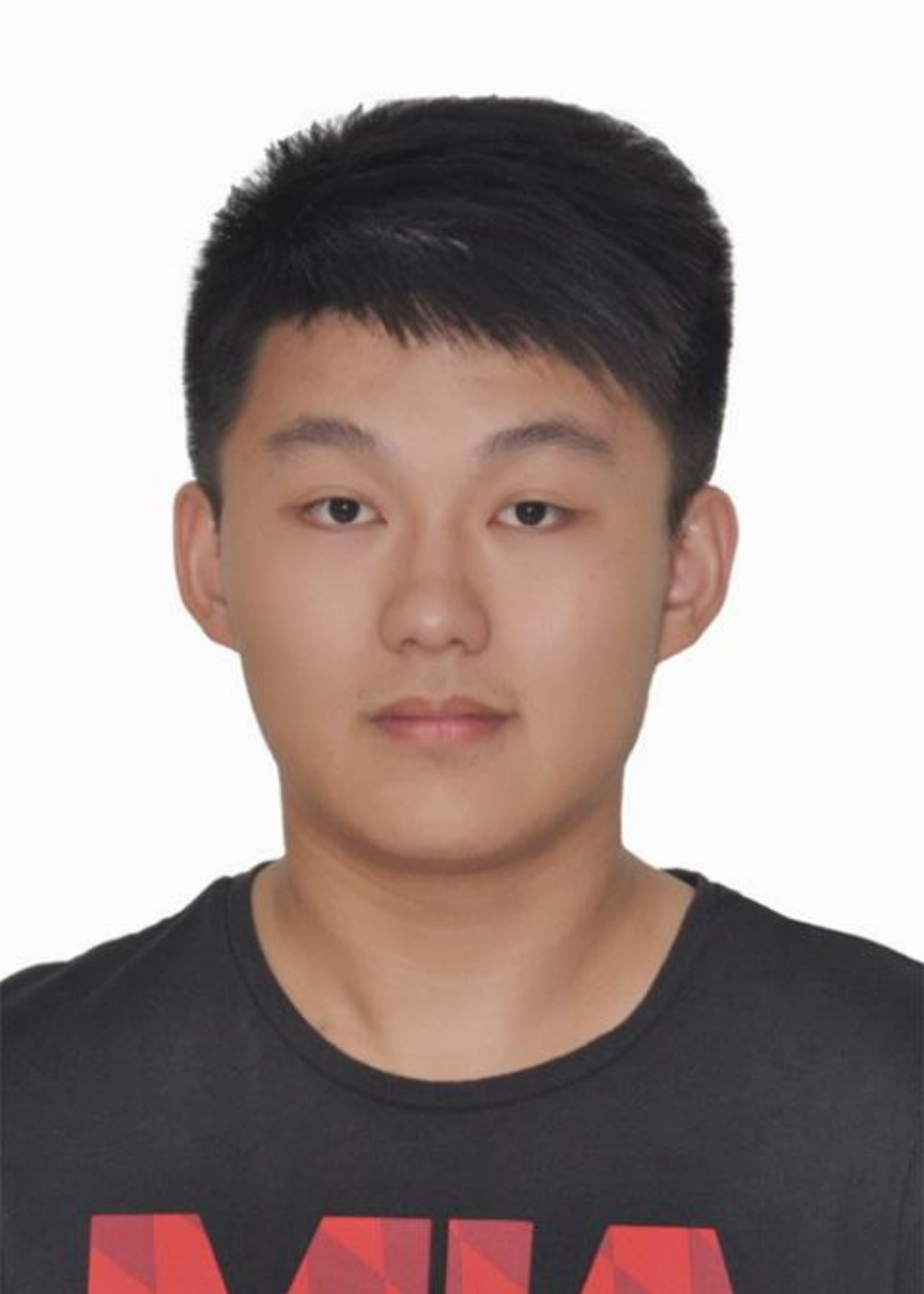}}]{Wentao Chen} is now working for an undergraduate degree at Beijing University of Posts and Telecommunications. His research interests include cyberspace security and deep learning.
\vspace{-12mm}
\end{IEEEbiography}

\begin{IEEEbiography}[{\includegraphics[width=1in]{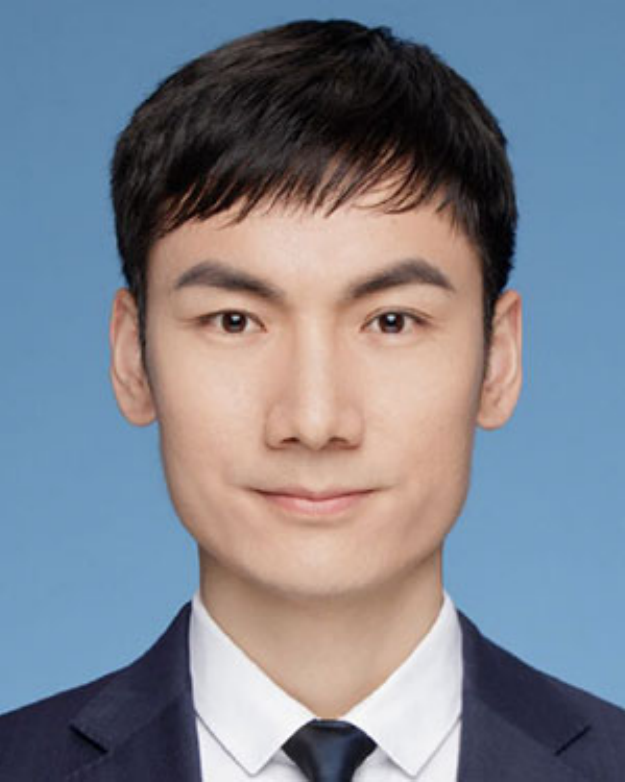}}]{Zhiqing Wei} (Member, IEEE) received the B.E. and Ph.D. degrees from the Beijing University of Posts and Telecommunications (BUPT) in 2010 and 2015. He is currently an Associate Professor with BUPT.
He has authored or coauthored one book, three book chapters and more than 50 articles. He was granted the Exemplary Reviewer of IEEE WIRELESS COMMUNICATIONS LETTERS in 2017, the Best Article Award of International Conference on Wireless Communications and Signal Processing (WCSP) 2018. He was the Registration Co-Chair of IEEE/CIC International Conference on Communications in China (ICCC) 2018 and the publication Co-Chair of IEEE/CIC ICCC 2019 and IEEE/CIC ICCC 2020. His research interest focuses on the performance analysis and optimization of machine type networks.
\vspace{-13mm}
\end{IEEEbiography}

\begin{IEEEbiography}[{\includegraphics[width=1in]{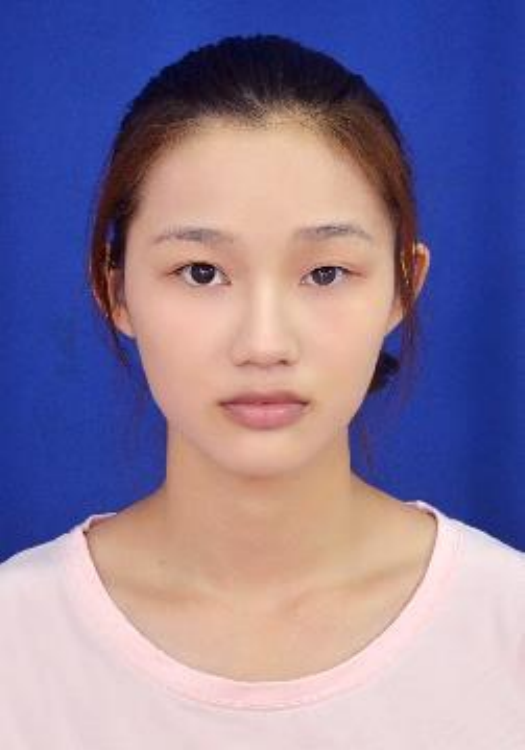}}]{Xingqi Luo} received the B.E. degree from Beijing Institute of Technology. Her research interests include cyberspace security.
\vspace{-13mm}
\end{IEEEbiography}

\begin{IEEEbiography}[{\includegraphics[width=1in]{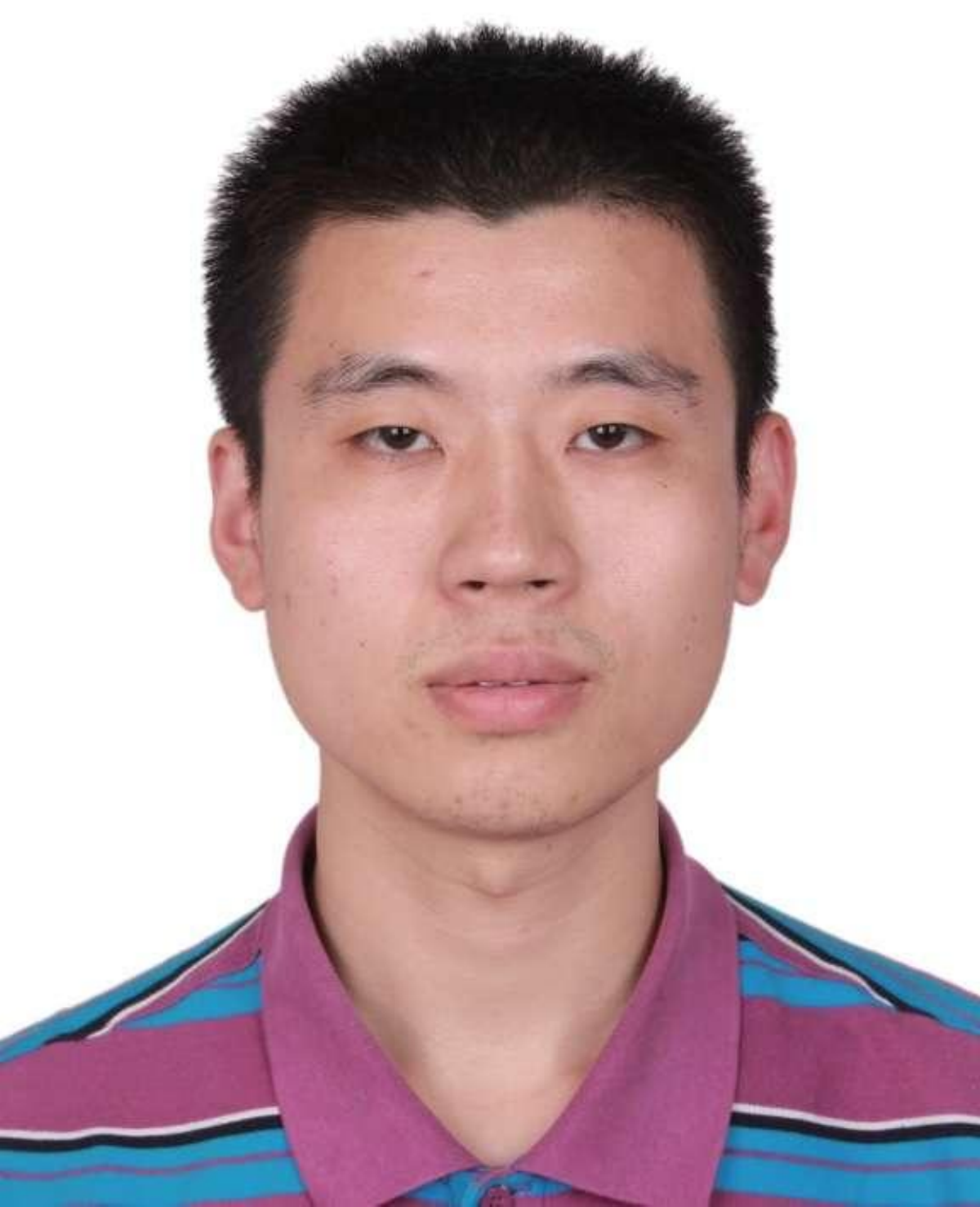}}]{Bing Su} received the BS degree in information engineering from the Beijing Institute of Technology, Beijing, China, in 2010, and the PhD degree in electronic engineering from Tsinghua University, Beijing, China, in 2016. From 2016 to 2020, he worked with the Institute of Software, Chinese Academy of Sciences, Beijing. Currently, he is an associate professor with the Gaoling School of Artificial Intelligence, Renmin University of China. His research interests include pattern recognition, computer vision, and machine learning. He has published more than ten papers in journals and conferences such as IEEE Transactions on Pattern Analysis and Machine Intelligence (TPAMI), IEEE Transactions on Image Processing (TIP), International Conference on Machine Learning (ICML), IEEE Conference on Computer Vision and Pattern Recognition (CVPR), IEEE International Conference on Computer Vision (ICCV), etc.
\vspace{-12mm}
\end{IEEEbiography}






\end{document}